\documentclass{pasj01}

\Received{$\langle$reception date$\rangle$}
\Accepted{$\langle$acception date$\rangle$}
\Published{$\langle$publication date$\rangle$}
\usepackage{lscape}
\usepackage{deluxetable}
\makeatletter
\def\@to{to}
\makeatother
\usepackage{multirow}

\begin{document}

\title{The formation of young massive clusters triggered by cloud-cloud collisions in the Antennae Galaxies NGC~4038/NGC~4039 }
\author{Kisetsu Tsuge\altaffilmark{1}, Yasuo Fukui\altaffilmark{1, 2}, Kengo Tachihara\altaffilmark{1},  Hidetoshi Sano\altaffilmark{1, 2}, Kazuki Tokuda\altaffilmark{3,4}, Junko Ueda\altaffilmark{5}, Daisuke Iono\altaffilmark{5,6}, Molly K. Finn\altaffilmark{7}}%
\altaffiltext{1}{Department of Physics, Nagoya University, Furo-cho, Chikusa-ku, Nagoya 464-8601, Japan}
\altaffiltext{2}{Institute for Advanced Research, Nagoya University, Furo-cho, Chikusa-ku, Nagoya 464-8601, Japan}
\altaffiltext{3}{Department of Physical Science, Graduate School of Science, Osaka Prefecture University, 1-1 Gakuen-cho, Naka-ku, Sakai, Osaka 599-8531, Japan}
\altaffiltext{4}{Chile Observatory, National Astronomical Observatory of Japan, National Institutes of Natural Science, 2-21-1 Osawa, Mitaka, Tokyo 181-8588, Japan}
\altaffiltext{5}{National Astronomical Observatory of Japan, 2-21-1 Osawa, Mitaka,Tokyo, 181-8588, Japan}
\altaffiltext{6}{SOKENDAI (The Graduate University for Advanced Studies), 2-21-1 Osawa, Mitaka, Tokyo 181-8588, Japan}
\altaffiltext{7}{Department of Astronomy, University Virginia, Charlottesville, VA 22904, USA}
\email{tsuge@a.phys.nagoya-u.ac.jp}


\maketitle

\begin{abstract}
The formation mechanism of super star clusters (SSCs), present-day analog{s} of the ancient globulars, still remains elusive. The major merger, the Antennae galaxies is forming SSCs and is one of the primary targets to test the cluster formation mechanism. We reanalyzed the archival ALMA CO data of the Antennae and found three typical observational signatures of a cloud-cloud collision toward SSC B1 and other SSCs in the overlap region; i. two velocity components with $\sim$100 km s$^{-1}$ velocity separation, ii. bridge features connecting the two components, and iii. {a} complementary spatial distribution between them, lending support {to} collisions of the two components as a cluster formation mechanism. We present a scenario that two clouds with 100 km s$^{-1}$ velocity separation collided, and SSCs having $\sim$10$^6$--10$^7$ $M_{\rm \odot}$ were formed rapidly during {that} time scale. {We compared the present results with the recent studies of star forming regions in the Milky Way and the LMC, where the SSCs having $\sim$10$^4$--10$^5$ $M_{\rm \odot}$ are located. As a result, we found that there is a positive correlation between the compressed gas pressure generated by collisions and the total stellar mass of SSC, suggesting that the pressure may be a key parameter in the SSC formation. }
\end{abstract}

\section{Introduction}
\subsection{Background}

\begin{figure*}[htbp]
\begin{center}
\includegraphics[width=10cm]{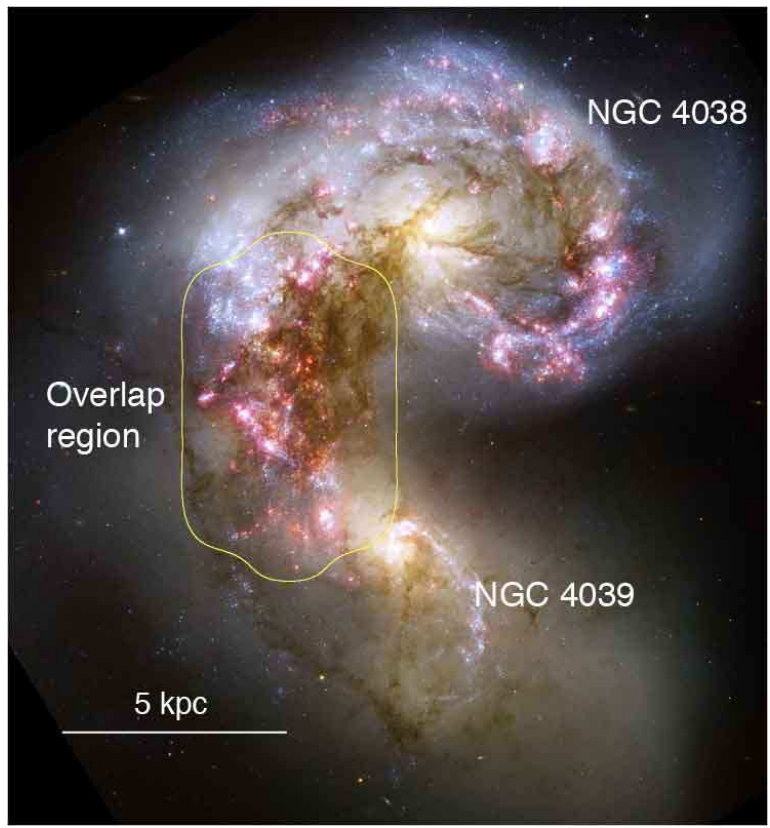}
\end{center}
\caption{Optical image of NGC~4038/4039 produced by data of the $Hubble$ $Space$ $Telescope$. B-band image is shown in blue, V-band image in green, and a combination of the I-band and H$\alpha$ images in red. The yellow contour shows the coverage of the ALMA {Cycle 0} observation.} 
\label{fig1}
\end{figure*}%

Super star clusters (SSCs) are rare objects in present-day normal galaxies and are {possibly young} counterparts to the globular clusters formed in the early Universe. In our efforts to understand galactic evolution, it is of primary importance to reveal the formation mechanism of SSCs in the local universe, which {could lead to insights into the formation and evolution of primordial galaxies.} It has been suggested that the interacting galaxies may be a site of SSC formation because the violent motion agitated by the galactic interaction may efficiently compress the gas into massive clusters (e.g., \cite{2000ApJ...542..120W}).

The Antennae galaxies, the closest major merger at a distance of 22 Mpc (\cite{2008AJ....136.1482S}), are interacting galaxies having rich molecular gas (\cite{2000ApJ...542..120W}), which may allow us to understand better the formation mechanism of SSCs. The definition of a SSC is not rigorous and different definitions are found in the literature. We adopt tentatively that a SSC is a stellar cluster whose mass is 10$^4$ $M_{\odot}$--10$^7$ $M_{\odot}$ within a volume of a 1 pc--10 pc radius. This definition is somewhat relaxed and includes {some spatially extended clusters, like NGC~6334 and NGC~604, with sizes of $\sim$10 pc, which are not listed as SSCs by} \citet{2010ARA&A..48..431P}.

The two spirals NGC~4038 (RA=$\timeform{12h01m53.0s}$, Dec=$\timeform{-18D52'10.0"}$) and NGC~4039 (RA=$\timeform{12h01m53.6s}$, Dec=$\timeform{-18D53'11.0"}$) are interacting {galaxies}. Numerical simulations of the galaxy encounter successfully reproduced the outcomes of the tidal interaction between the two galaxies, including two long tails, clear signatures of tidal interaction, and the overlap region with heavy obscuration (\cite{1993ApJ...418...82M,2004MNRAS.350..798B,2007A&A...468...61D,2008MNRAS.389L...8B,2008MNRAS.384..386C,2009ApJ...706...67R,2009PASJ...61..481S,2010ApJ...715L..88K,2010ApJ...720L.149T,2013MNRAS.434.1028P}). The interaction has been continuing probably since a few 100 Myr ago and the two spirals may eventually merge together into a larger galaxy (\cite{2015MNRAS.446.2038R}). The unusually active star formation including {thousands} of SSCs and their candidates is an outstanding aspect (e.g., \cite{1995AJ....109..960W}). The near infrared imaging data taken with $Hubble$ $Space$ $Telescope$ (\cite{2007ApJ...668..168G,2010AJ....140...75W}) are used to {identify} SSCs. The two galaxies are apparently merging in the overlap region, where five out of eight SSCs have {total stellar} masses larger than 10$^6$ $M_{\odot}$ as listed in Table 1. Among all, the most luminous SSC B1 in the Antennae was identified by \citet{1995AJ....109..960W}. \citet{2000ApJ...542..120W} mapped the CO emission and found that the CO clouds are organized into several complexes as named super giant molecular complexes (SGMCs) 1--5. These authors suggested a possibility of collision between SGMCs as origin of strong mid-infrared emission, but its relationship with the SSC formation was not discussed further.

\begin{deluxetable}{ccccc}
\tablewidth{6.0cm}
\tablecaption{Young massive SSCs in  southern part of the overlap region}
\label{tab:ssc}
\tablehead{\multicolumn{1}{c}{Object} &Age&{$M_{\rm cluster}$}   & Ref.  \\
\multicolumn{1}{c}{\ }&{[}Myr{]} &{[}10$^{6}$$M_{\odot}${]}&\  \\
\multicolumn{1}{c}{(1)} &(2)&(3)&(4)}
\startdata
D & 3.9& 1.4 & {[}1{]} \\
\  & 1.45& 0.32&  {[}2{]} \\ \hline
D1 & 6.1& 1.6&  {[}1{]} \\  \hline
D2 & 5.4& 0.8& {[}1{]}  \\  \hline
C & 5.7& 4.1&  {[}1{]} \\ 
\  & 4.8& 1.2&  {[}2{]} \\ \hline
B1 & 3.5& 4.2&  {[}1{]} \\ 
\  & 1& 6.8&  {[}2{]} \\ \hline
\enddata
\tablecomments{Column (1): Object name. Column (2): Age of object. Column (3): Total stellar mass of the cluster. Column (4): [1] Gilbert \& Graham (2007),  [2] Whitmore et al. (2010).}
\end{deluxetable}

The ALMA CO data offer an ideal opportunity to investigate the molecular gas distribution and kinematics which are likely connected to the cluster formation and feedback at resolution better than 10 pc. The previous works on the ALMA data investigated the overall filamentary distribution of the molecular clouds (\cite{2014ApJ...795..156W}), a candidate proto-cluster dubbed “{the} Firecracker” in the overlap region (\cite{2012A&A...538L...9H,2015ApJ...806...35J}), and the high pressure environment in the overlap region (\cite{2015ApJ...806...35J}). \citet{2014ApJ...795..156W} discussed a possible role of filamentary cloud collision in triggering cluster formation toward {the} Firecracker (see also \cite{2012A&A...538L...9H}) {by using the Cycle 0 data}. The molecular gas motion around SSC~B1 was interpreted in terms of a mixture of the parent cloud of the cluster and a cloud unrelated to the cluster formation, and it was suggested that part of the gas shows expansion due to feedback (\cite{2017A&A...600A.139H}). Most recently, \citet{2019ApJ...874..120F} presented the highest resolution {image of “{the} Firecracker” with ALMA, and discussed that it is a promising candidate of a proto-SSC.}

{We present the reanalyzed data of the ALMA Cycle 0 data in the followings. Figure 1 shows the optical image of the Antennae Galaxies and the distribution of the $^{12}$CO (3--2) emission obtained with ALMA Cycle 0 observations in the overlap region, while Figure 2 shows an enlarged view of the southern part.}

\begin{figure}[htbp]
\begin{center}
\includegraphics[width=\linewidth]{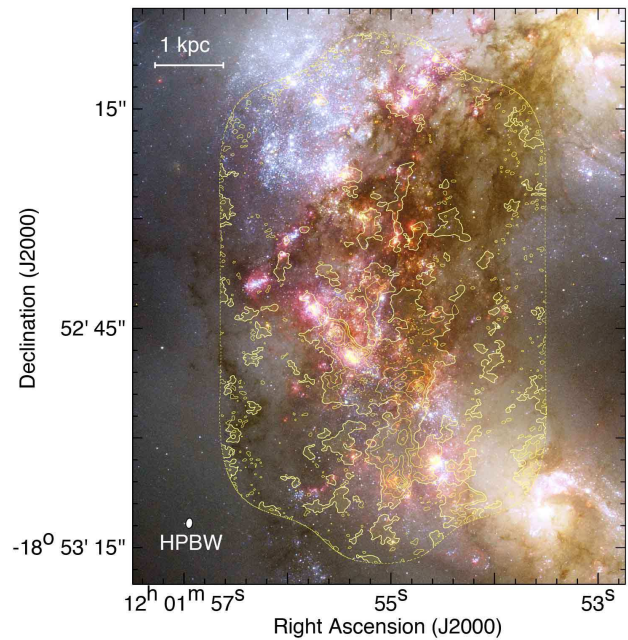}
\end{center}
\caption{Total integrated intensity map of $^{12}$CO (3--2) toward the overlap region of the Antennae Galaxies superposed on the same $HST$ image as shown in Figure 1. {The synthesized beam size of CO is {0$\farcs$70}$\times${0$\farcs$46}.} The contour levels are 22.9 (3$\sigma$), 100, 300, 500, 700, and 900 K km s$^{-1}$. The integration velocity range is $V$$_{\rm LSR}$ = 1300--1795 km s$^{-1}$.} 
\label{fig2}
\end{figure}%

\subsection{The triggered formation of SSCs}
It is argued that the high pressure of the cluster formation environment is a requirement for the SSC formation (e.g., \cite{1997ApJ...480..235E}). The “high pressure” alone is not a complete description of the formation mechanism because the cloud dynamics is not explicitly given, leaving the actual mechanism ambiguous. Numerical simulations aimed at the SSC formation show that the collection of a large gas mass into a small volume is difficult (Inoue et al. 2019). In the course of the collecting process the gas density becomes too high to avoid forming stars before forming a compact cloud core. The {collection} process must be rapid enough so that the gas becomes confined within 1--10 pc without being consumed by extended non-clustered star formation.

{Recent papers on triggered high-mass star formation by cloud-cloud collision raise a possibility to apply cloud-cloud collision to the Antennae. More than 50 O-star clusters/single O stars were identified as formed by cloud-cloud collision until now in the literature including the cloud-cloud collision special issue of PASJ in 2018. In these works, two empirical signatures of collision, i.e., the complementary spatial distribution between the two clouds and/or the bridge feature in a velocity space between them, are employed. These two signatures are formed by collision, because the clouds disrupt each other by collisional compression and cause thereby momentum exchange which shifts the interacting gas into a wide velocity range between the initial two cloud velocities. In order to base the signatures on a theoretical basis, Fukui et al. (2018a) made synthetic observations of cloud-cloud collision by employing the numerical simulations of collision of Takahira et al. (2014), and confirmed these two observable signatures {as viable}. In addition, Fukui et al. (2018a) showed that the complementary distribution usually accompanies a displacement between the two clouds due to the projection effects and showed that the method successfully applied to the cloud-cloud collisions in {M43}; if the collision takes place along the line of sight, the two colliding clouds show no displacement in the sky.} {The positions of two clouds shift in the sky after collision, if the collision path has a certain finite angle to the line of sight, resulting an apparent displacement in the complementary distribution.} {See Fukui et al. (2018a) for more details of }{these collision signatures} {and also \citet{2015MNRAS.450...10H} for additional properties of the bridge features. }


\begin{figure}[htbp]
\begin{center}
\includegraphics[width=\linewidth]{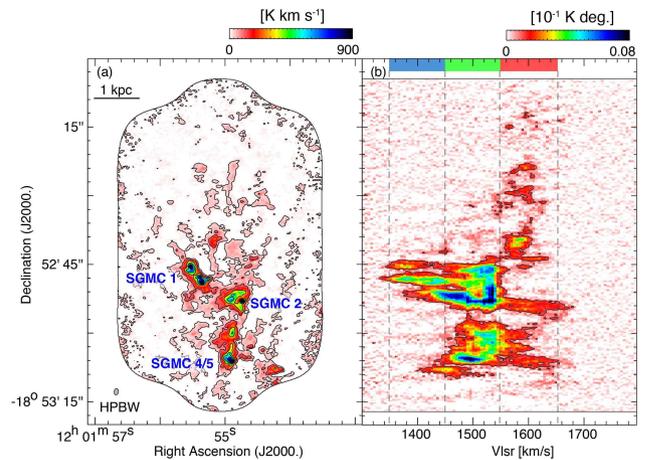}
\end{center}
\caption{ (a) Total integrated intensity map of $^{12}$CO (3--2) toward the overlap region. The contour levels and integration velocity range are the same as in Figure \ref{fig2}. (b) Dec.--velocity diagram of $^{12}$CO (3--2) . The integration range in R.A. is the whole region of (a). The contour levels is 10$\sigma$ (0.001 K deg.). {The vertical dashed lines and blue, green, and red color bars indicate the velocity ranges of the blue-, green-, and red-clouds.}} 
\label{fig3}
\end{figure}%

In the Milky Way, {\citet{2010ARA&A..48..431P} report more than 10 known SSCs with masses $>$10$^4$ $M_{\odot}$} in addition to another cluster RCW~38 which was later identified as a SSC (\cite{2015ApJ...802...60K}; see also \cite{2016ApJ...820...26F}). {The associated molecular gas {toward} the five youngest SSCs with ages less than $\sim$2 Myr were studied in detail, and cloud-cloud collision {is} identified as {a} trigger of each cluster's formation (Westerlund~2, \cite{2009ApJ...696L.115F}; NGC~3603, \cite{2014ApJ...780...36F}; RCW~38, \cite{2016ApJ...820...26F}; DBS[2003]179, \cite{kuwahara2019inprep}; {Trumpler}~14, \cite{fujita2019ainprep}).} 
The remaining older clusters with an ages more than {$\sim$3} Myrs have little molecular gas due to ionization within 10 pc of the cluster, and it is possible that the collision signatures are lost even {if} their formation was triggered by a cloud-cloud collision.

In the LMC, R136 having 10$^5$ $M_{\odot}$ is a SSC, where H{\sc i} colliding flows at $\sim$50 km s$^{-1}$ {likely} triggered {the cluster} formation as revealed by an H{\sc i} study of \citet{2017PASJ...69L...5F}. The high velocity flow was induced by the tidal interaction between the LMC and SMC 0.2 Gyrs ago, and lead to the recent collision $\sim$1 Myr ago after orbital motion around the LMC (\cite{1990PASJ...42..505F,2007PASA...24...21B,2014MNRAS.443..522Y}). {Recent ALMA observations toward the N159 GMC, which is located at $\sim$500 pc south of R136, found high-mass star-forming CO filaments (\cite{2015ApJ...807L...4F,2017ApJ...835..108S}) with the typical width of $\sim$0.1 pc {(\cite{2019ApJ...886...14F,2019ApJ...886...15T})}. These results suggested that high-mass star formation activities in the molecular ridge (i.e., Southeast part of the LMC) {are triggered by the same H{\sc i} flow over a kpc scale} (\cite{2017PASJ...69L...5F,2019ApJ...871...44T}).}
Subsequently, it was shown that two O star forming clusters, NGC~604 in M33 having $\sim$10$^5$ $M_{\odot}$ and NGC~602 in the SMC having at least 5500 $M_{\odot}$, exhibit a signature of H{\sc i} colliding flows (\cite{2018PASJ...70S..52T,sano2019inprep}). These results suggest that the supersonic collision between two H{\sc i} flows is a usual mechanism for the formation of SSCs of 10$^4$ $M_{\odot}$--10$^5$ $M_{\odot}$ in the Local Group. It seems therefore natural that a cloud-cloud collision, if the compression is strong enough, leads to the formation of a more massive cluster similar to the globulars, and we are able to test the possibility in the Antennae by using the collision signatures above. We aim to carry this out in the present paper. Although the converging H{\sc i} flows in the Local Group are not of extremely high density, the highly supersonic collision is very powerful in compressing the gas to molecular gas in a short time scale within a few Myrs as demonstrated by MHD simulations of converging gas flows (\cite{2012ApJ...759...35I,2013ApJ...774L..31I,2018PASJ...70S..53I}).

This paper is organized as follows. Section 2 summarizes the ALMA archive data reanalyzed in the present work, and Section 3 shows the results, the CO distributions at $\sim$0.6 arcsec resolution of Band 7. Sections 4 and 5 {discuss the details of the cluster/candidate cluster formation} and Section 6 summarizes the paper.

\begin{figure}[htbp]
\begin{center}
\includegraphics[width=\linewidth]{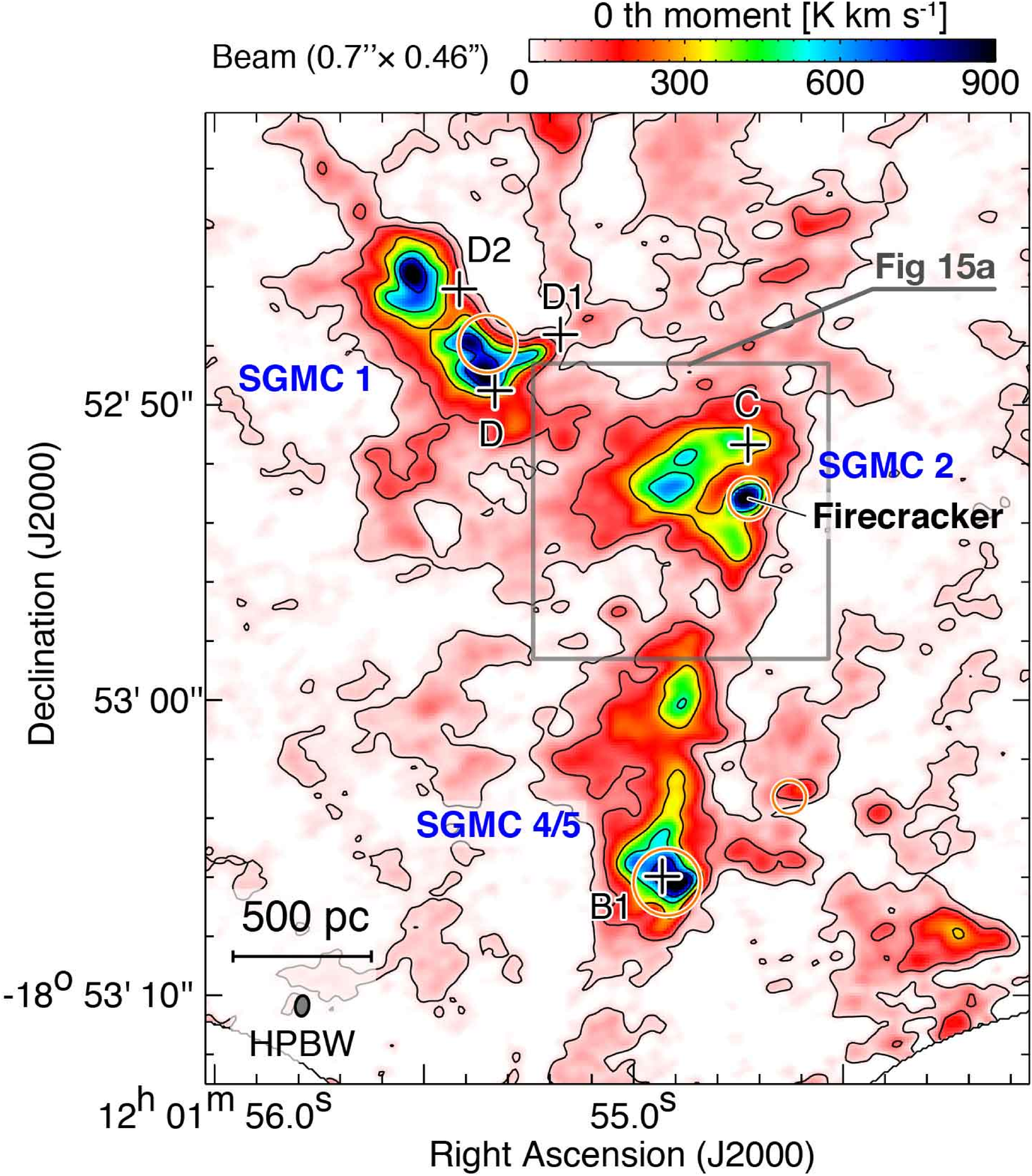}
\end{center}
\caption{Total integrated intensity map of $^{12}$CO (3--2) toward south part of the overlap region. The contour levels and integration velocity range are the same as in Figure \ref{fig2}. The black crosses and orange circles indicate the positions of superstar clusters (\cite{2007ApJ...668..168G}) and H$_{2}$ emission source (\cite{2012A&A...538L...9H}), respectively.} 
\label{fig4}
\end{figure}%

\section{The ALMA archive of the Antennae and data reduction}
\subsection{Cycle 0 data toward the overlap region: GMC scale}
In the present work we made use of the archival $^{12}$CO (3--2) data of Band 7 (345 GHz) of the Antennae (NGC~4038/{40}39) from Cycle 0 project {(\#2011.0.00876, P.I., B. Whitmore)}. {Whitmore et al. (2014) used a briggs weighting for the imaging, and reultant synthesized beam size is  {0$\farcs$56}$\times${0$\farcs$43}.} The detailed descriptions of the observations are given by \citet{2014ApJ...795..156W}. We re-performed the imaging process from the visibility data by using the CASA (Common Astronomy Software Application) package (\cite{2007ASPC..376..127M}) version 5.0.0. We used the multiscale CLEAN algorithm (\cite{2008ISTSP...2..793C}) implemented in the CASA {with natural weighting}. The resultant synthesized beam size is {0$\farcs$70}$\times${0$\farcs$46}, and the 1$\sigma$ RMS noise is $\sim$4.8$\times$10$^{-3}$ Jy beam$^{-1}$ at a velocity resolution of 5.0 km s$^{-1}$, which corresponds to a surface brightness sensitivity of $\sim$0.15 K. {We recovered extended emission and reduced} {negative levels} {by using multiscale CLEAN algorithm {(see Figure 2)}.} {Total flux is recovered $\sim$54\%, and hence the extended components are important.} {D}{etails of} {a} {comparison with {the }data used in Whitmore et al. (2014) are shown in Appendix 1.}
{T}o investigate the missing flux, we compared the peak brightness temperature of the single-dish (JCMT) data (\cite{2003ApJ...588..243Z}) and our re-processed data smoothed to the single-dish resolution {of} $\timeform{14"}$. {Although the ALMA flux is $\sim$20\% smaller than the single-dish data, the relative error in mass estimation is 20--50\% with an average of 30\%, as shown in Table 2.} Therefore, the missing flux is {smaller than the uncertainty of mass estimation and} considered to be not significant. 

\begin{figure*}[htbp]
\begin{center}
\includegraphics[width=\linewidth]{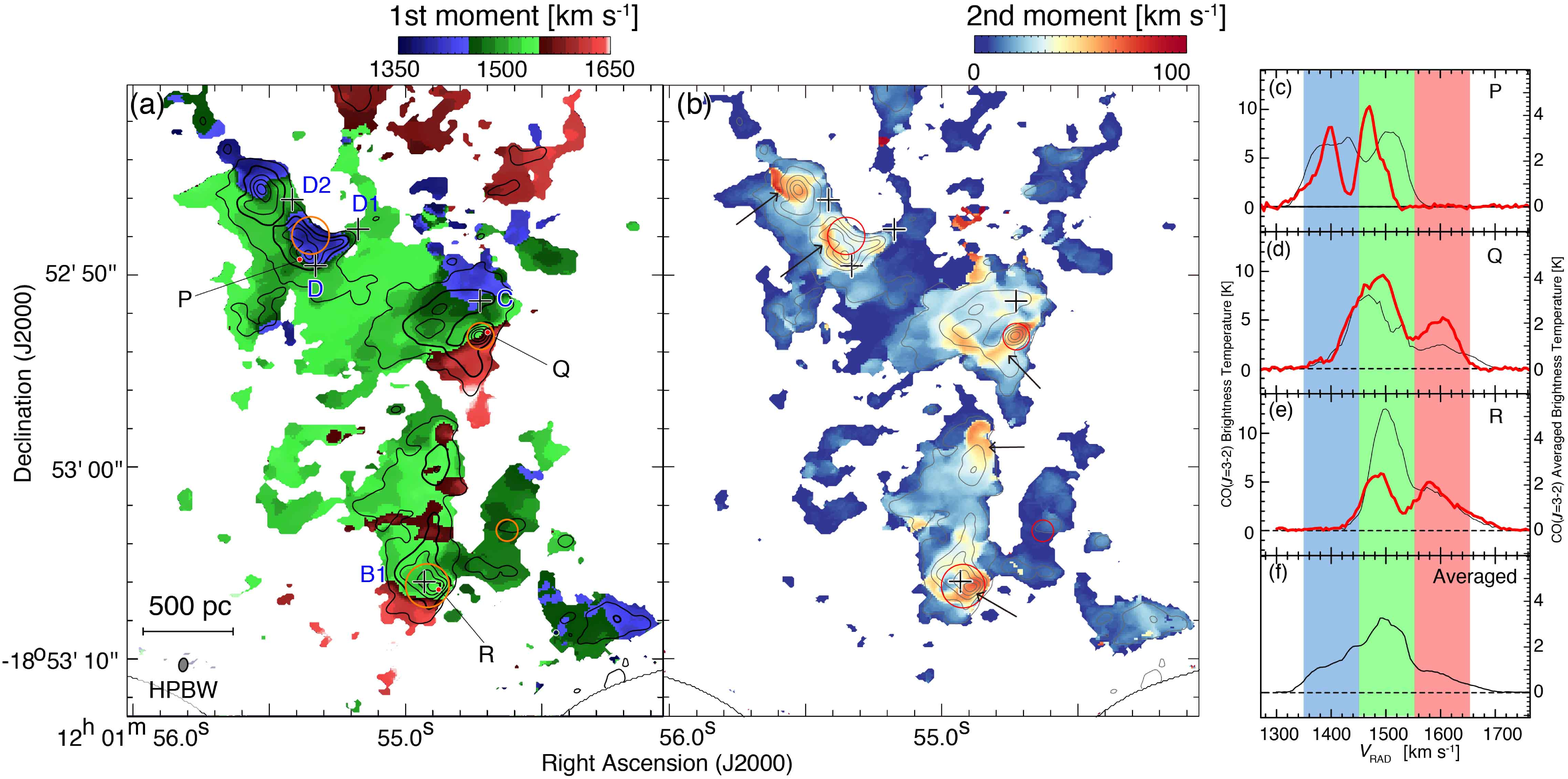}
\end{center}
\caption{(a) 1st moment map of $^{12}$CO (3--2) toward south part of the overlap region derived using miriad task. 1st moment {is the intensity-weighted velocity following the equation of} $\Sigma(I\times v)/\Sigma(I)$, where $I$ is intensity of emission and $v$ is velocity. The superposed contours indicate the {integrated intensity map} of $^{12}$CO (3--2) . The lowest contour level and intervals are 100 and 200 K km s$^{-1}$. The integration velocity range is the same as in Figure 2. (b) 2nd moment map of $^{12}$CO (3--2) toward south part of the overlap region. 2nd moment {is the velocity dispersion following the equation of} $\sqrt{\Sigma(I\times(v-M1)^{2})/\Sigma(I)}$, where M1 is the 1st moment. The black crosses are the same as in Figure 4. {Values larger than 1.5 K (1$\sigma$) are used for making 1st and 2nd moment maps.} {Red profiles in} (c), (d), and (e) are typical spectra of $^{12}$CO (3--2) toward integrated intensity peaks of SGMCs at the positions of P (R.A., Dec. = $\timeform{12h01m55.39s}$,$\timeform{-18D52'49.2"}$), Q (R.A., Dec. = $\timeform{12h01m54.7s}$,$\timeform{-18D52'53.0"}$), and R (R.A., Dec. = $\timeform{12h01m54.88000s}$,$\timeform{-18D53'06.4"}$) in (a), respectively. {Black profiles in (c), (d), and (e) are averaged profiles toward SGMC~1, SGMC~2, and SGMC~4--5, respectively.} Blue, {green}, and red shaded ranges are $V$$_{\rm LSR}$ = 1350 to 1450 km s$^{-1}$ ({blue-cloud}), 1450 to 1550 km s$^{-1}$ ({green-cloud}), and 1550 to 1650 km s$^{-1}$ ({red-cloud}), respectively. {(f) Averaged profile of $^{12}$CO (3--2) toward the southern part of overlap region shown in Figure 5a. The regions averaged the CO spectra of Figures 5c--5f are inside of the second lowest contour of Figure 5a.}} 
\label{fig5}
\end{figure*}%

\subsection{Cycle 4 data toward the {Firecracker}: cluster scale}
We also used the archival $^{12}$CO (3--2) data of Band 7 {toward {the} Firecracker in SGMC2} from {ALMA} Cycle 4 project {(\#2016.1.00924.S, P.I., K. Johnson)} {and combined with the Cycle 0 data}. The detailed descriptions of the {Cycle 4} observations are given by \citet{2019ApJ...874..120F}. {In order to make the analysis consistent with the other regions, we redid the imaging process from the visibility data by using the CASA package version 5.4.1. and the multiscale CLEAN algorithm same as for the Cycle 0 data (Section 2.1.).} The resultant synthesized beam size is $\sim$0$\farcs$16$\times$0$\farcs$19 {for natural weighting.} The 1$\sigma$ RMS noise is $\sim$1.3$\times$10$^{-3}$ Jy beam$^{-1}$ at a velocity resolution of 5.0 km s$^{-1}$, which corresponds to a surface brightness sensitivity of $\sim$0.45 K. {The result is consistent with that given by Finn et al. (2019) who used briggs weighting.}

\section{Results of the Reanalysis}
\subsection{Intensity Distribution}

Figure 3a shows the distribution of the $^{12}$CO (3--2) emission obtained with ALMA in the overlap region of the Antennae. The three regions with clusters are denoted as SGMC1, SGMC2, and SGMC4--5 (\cite{2014ApJ...795..156W,2000ApJ...542..120W}). Figure 3b shows a Declination-velocity diagram of the overlap region, which indicates that the southern part at declination less than $\timeform{-18D52'40.0"}$ shows particularly broad CO emission extended over $\sim$300 km s$^{-1}$ as compared with the northern part with a velocity span less that $\sim$100 km s$^{-1}$. {In Figure 3b we show the three velocity ranges with around 100 km s$^{-1}$ spans, which are divided based on the velocity field in the molecular clouds as explained in Section} {3.3.}. {Figure 4 shows an enlarged view of the southern part of Figure 3.}
We {selected five clusters having ages of $\leq$6 Myr and {total stellar} masses of $\geq$10$^6$ $M_{\odot}$ in the overlapping area listed in Table 1 from the cluster catalog (\cite{2014ApJ...795..156W,2007ApJ...668..168G}). The five clusters as well as {the} Firecracker are all located toward the three cloud complexes (Figure 4). }


\begin{figure}[htbp]
\begin{center}
\includegraphics[width=\linewidth]{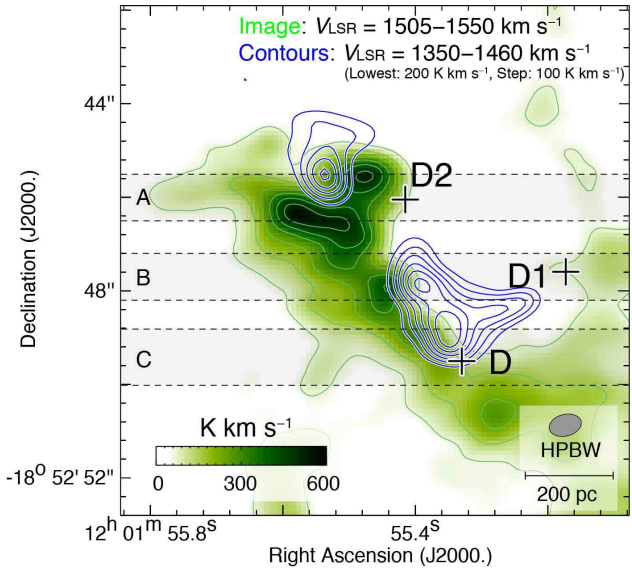}
\end{center}
\caption{The $^{12}$CO (3--2) intensity map of {blue-}cloud by {blue} contours superposed on the {green-}cloud toward SGMC 1. The integration velocity range is $V$$_{\rm LSR}$ = 1490--1550 km s$^{-1}$ for the green-cloud and $V$$_{\rm LSR}$= 1350--1450 km s$^{-1}$ for the blue-cloud. The lowest contour level and intervals are 150 and 50 K km s$^{-1}$, respectively. The black crosses indicate the positions of SSCs D, D1, and D2 (\cite{2007ApJ...668..168G}). {The red crosses indicate the positions of the embedded clusters (Whitmore et al. 2014).} The black shaded regions show the integration range in R. A. in Figure \ref{fig7}. }
\label{fig6}
\end{figure}%

\begin{figure*}[htbp]
\begin{center}
\includegraphics[width=\linewidth]{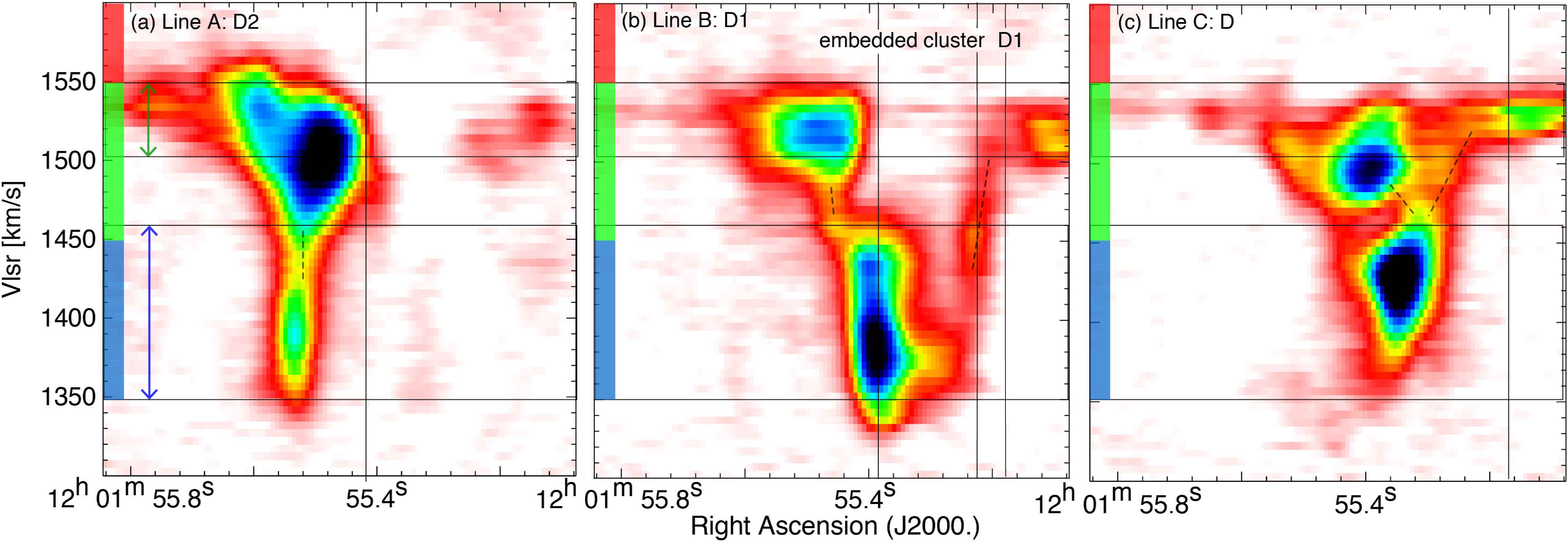}
\end{center}
\caption{Declination--velocity diagram of $^{12}$CO (3--2) along the lines A, B, and C shown in Figure 6. The {black horizontal lines} indicate the velocity ranges of the blue- and green-clouds. {The black vertical lines indicate the R.A. of the positions of the {embedded clusters and} SSCs. The black dashed lines show the bridge features in the intermediate velocity range between the blue- and green-clouds.}}
\label{fig7new}
\end{figure*}%

Figure 5 shows the distributions of the 1st and 2nd moment along with the {$^{12}$CO (3--2)} line profiles toward {the} {peak of the} three {SGMCs} named P, Q, and R {(red line)}{, where the total integrated profile (black line) is overlaid. }

{The variation of the velocity structures are shown in {the} 1st moment map and the position-velocity diagrams for the whole cloud in Appendix 2.} In Figure 5a, the 1st moment shows that the velocity is often discontinuous with a velocity jump of $\sim$100 km s$^{-1}$ toward the CO peaks. {The clouds mainly consist of three discrete velocity components, referred to as blue- (1350 km s$^{-1}$), green- (1450 km s$^{-1}$), and red-clouds (1550 km s$^{-1}$).} The 2nd moment is enhanced in five filamentary regions as indicated by arrows in Figure 5b. We see a trend that the filamentary regions are distributed along the velocity discontinuity in Figure 5a. {Each {peak} profile in Figures 5c, 5d, and 5e shows two peaks in a range from} 1400 km s$^{-1}$ to 1700 km s$^{-1}$, which are separated in velocity by 80--110 km s$^{-1}$. {The peak velocities are roughly consistent with the three velocities above.} 
{Two velocity components are also seen in averaged spectrum of each SGMC, and these are consistent with the typical spectrum at the integrated intensity peaks of SGMCs. Figure 5f is the averaged spectrum of the southern part of the overlap region. The green-cloud has the highest intensity and is possibly considered to be the main velocity component among three velocity components.}
{Specifically, we identified the velocity components in each SGMC. By using the 1st and 2nd moments in Figure 5 {and channel maps of declination--velocity diagrams in Figure{s} 20, 21, and 22}, we defined the two velocity ranges in each SGMC as follows;} {the ranges are 1350--1460 km s$^{-1}$ and 1505  --1550 km s$^{-1}$ for SGMC1, 1400--1530 km s$^{-1}$ and 1580--1700 km s$^{-1}$ for SGMC2, and 1450--1535 km s$^{-1}$ and 1560--1700 km s$^{-1}$ for SGMC4--5.} {The SGMCs spread over several hundreds of pc, and each velocity of the two components shifts with positions inside of the SGMCs. The velocity ranges of the bridge features are thus defined as the velocity where the bridge features exist at any place in each SGMC.}

\subsection{{Physical properties of SGMCs}}
 {In order to derive physical parameters of the clouds,} { first, we performed {spectral line} fitting to the Gaussian profile at the position where the integrated intensity is maximum for each cloud, and derived peak velocity, peak intensity, and velocity dispersion. The integral velocity ranges are shown in the last paragraph of Section 3.1.}  {We calculated the cloud mass of the region enclosed by a contour of 20\% of {the} peak integrated intensity ($\sim$30 $\sigma$). At} {a} {lower level like} 5 $\sigma$, {SGMCs are connected} {with each other }{by diffuse components as shown in Figure 4 and cannot be separated.}

{The virial masses are derived from the virial theorem by using the velocity dispersion 
($\Delta V$) and the equivalent radius of CO clouds ($r$) assuming spherical shapes. We use the equation by assuming a uniform density sphere of radius $r$ as follows:}
{
\begin{eqnarray}
M_{\rm vir} / M_{\rm \odot} = 210  \it r  (\rm d \it V)^2 M_{\rm \odot}
\end{eqnarray}
}

{$X_{\rm CO}$ mass was estimated by using the CO-to-H$_{2}$ conversion factor $X_{\rm CO}$= 0.6$\times$10$^{20}$ cm$^{-2}$ (K km s$^{-1}$)$^{-1}$(\cite{2003ApJ...588..243Z,2014ApJ...795..174K}). {The value of $X_{\rm CO}$ possibly changes by a factor of three because of the difference of the relative abundance of CO to H$_{2}$.} We use the equation as follows: 
\begin{eqnarray}
N (\rm H_{2}) = \it X_{\rm CO} \times \it{W}_{\rm CO},
\end{eqnarray}
where $W_{\rm{CO}}$ is the {integrated} intensity of the $^{\rm 12}\rm{CO}(\it{J}\rm{ = 1-0})$ and $N$(\rm H$_{2}$) is the column density of molecular hydrogen.} 

{As Ueda at al. (2012) showed that the ratio {of CO(3--2)/(1--0)} in the southern part of the overlap region is up to 0.3--0.5, we adopted the $^{12}$CO (3--2) to $^{\rm 12}$CO (1--0) ratios of SGMC~1, SGMC~2, SGMC~4--5, {and the Firecracker} are 0.4$\pm$0.1, 0.3$\pm$0.1, 0.5$\pm$0.1, and 0.3$\pm$0.1, respectively.} {We define} {the} {cloud size as an effective radius of ($A$/$\pi$)$^{0.5}$, where $A$ is the region enclosed by a contour of 20\% of peak integrated intensity.} The molecular mass of each CO cloud ranges from $\sim$3$\times$10$^7$ to 2$\times$10$^8$ $M_{\odot}$ with a typical size of 150--250 pc. Table 2 lists the physical parameters of the {individual} clouds.
Further, we estimate{d} the total molecular mass {of the whole overlap region enclosed by a contour of 3$\sigma$ level of Figure 2} to be 1.1$\times$10$^9$ $M_{\odot}$, which amounts to 3 times the total mass of the three CO clouds in Table 2. {The mass of the enveloping gas is therefore dominant in the overlap region.}

\subsection{Detailed velocity distribution}
{In order to better understand the cloud kinematics, we pursue the relationship between the multiple velocity components (Section 3.2) into more detail.} Figure 6 shows an overlay of the two velocity components in SGMC1 including D, D1, and D2. 
{In Figure \ref{fig7new}, R.A.--velocity diagrams toward SSC D2, D1, and D are shown for a strip denoted by the black shaded regions in Figure \ref{fig6}. }Figure \ref{fig6} presents that the two velocity components show a complementary distribution in the sense that the two components are distributed exclusively with each other; the blue-cloud has two peaks and the {green-cloud} is located between them along the southeastern edge of the blue-cloud.
{Significant bridge features toward {embedded clusters,} D2, D1, and D are seen at RA=$\timeform{12h1m55.52s}$ (Figure 7a), $\timeform{12h1m55.42s}$ and $\timeform{12h1m55.2s}$(Figure 7b), and $\timeform{12h1m55.3s}$(Figure 7c), respectively.}

\begin{figure*}[htbp]
\begin{center}
\includegraphics[width=\linewidth]{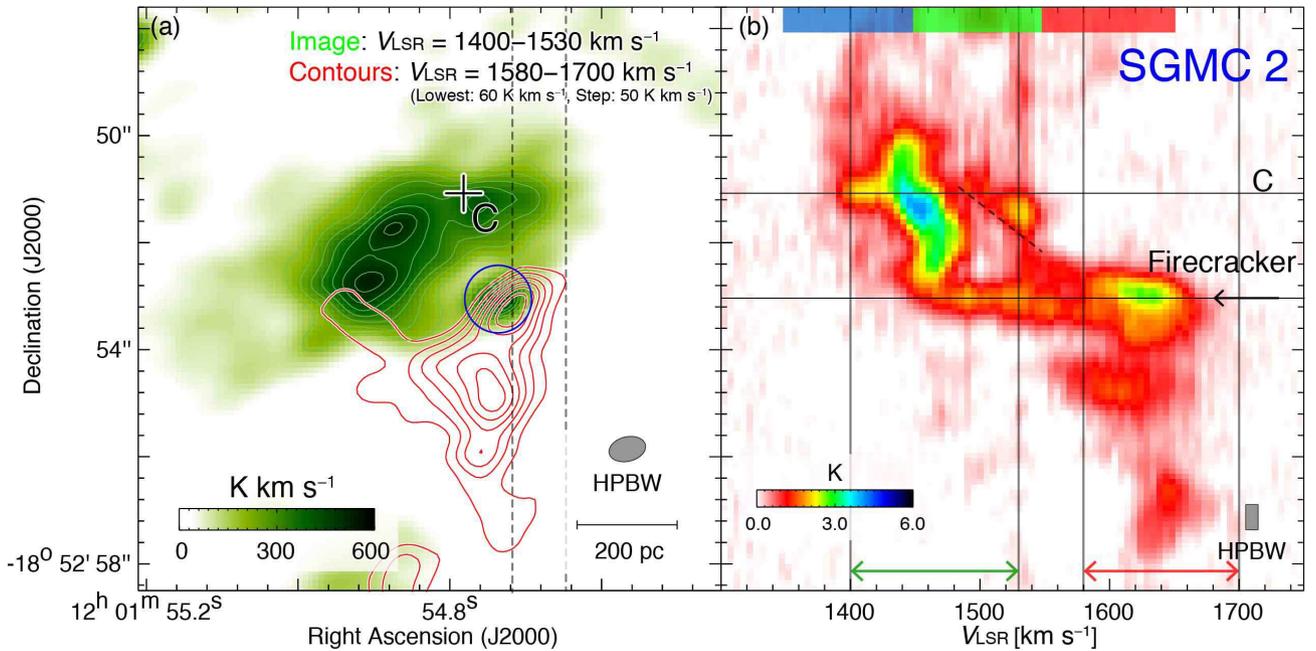}
\end{center}
\caption{(a) $^{12}$CO (3--2) intensity map of red-cloud ({red} contours) superposed on the {green-}cloud {green image and contours} toward SGMC 2. The integration velocity range is $V$$_{\rm LSR}$ = 1570--1700 km s$^{-1}$ for the red-cloud and $V$$_{\rm LSR}$ = 1400--1520 km s$^{-1}$ for the green-cloud. The lowest contour level and intervals are 60 and 50 K km s$^{-1}$, respectively. The black cross indicates the position of SSC C (\cite{2007ApJ...668..168G}). The black vertical lines show the integration range in R. A. in (b). (b) Declination-velocity diagram of $^{12}$CO (3--2). The {black horizontal lines} indicate the positions of the SSC C and {the} Firecracker.  The black dashed lines show the bridge feature in the intermediate velocity range between the green- and red-clouds.} 
\label{fig7}
\end{figure*}

Figure \ref{fig7}a shows an overlay of the two velocity components in SGMC2. We see the two velocity components show complementary distribution. Figure \ref{fig7}b shows a bridge feature connecting the two components at 1530--1580 km s$^{-1}$ in a Declination-velocity diagram. {This feature is already noted in the previous works (Whitmore et al. 2014; Finn et al. 2019).}

Figure \ref{fig8}a shows the two velocity components for SGMC 4--5 in a similar manner as in SGMC 1 and 2, also indicating complementary distribution as detailed in {Section 4}. We recognize {a hint of} a small displacement of $\sim$100 pc between them. Figure \ref{fig8}b is a position-velocity diagram which indicates {two} bridge features {at} declinations of $\timeform{-18D53'6.0"}$ {and} $\timeform{-18D53'4.0"}$ {as shown by black {dashed lines}}.


In addition to the two components, we find the bridge features in each of the SGMC{s} in the intermediate velocity range. Figure \ref{fig9} shows the spatial distributions of the bridges {($V_{\rm LSR}$=1460--1505 km s$^{-1}$)} superposed on the two velocity components in SGMC1. The bridges are distributed so as to {fill the gap between} the two velocity components near the SSCs{, in particular toward D.} Figure \ref{fig10} shows the spatial distributions of the bridges {($V_{\rm LSR}$=1530--1580 km s$^{-1}$)} superposed on the two velocity components in SGMC2. The bridges are {largely} located at the boundary of the two velocity components. In particular, the bridges have a peak of integrated intensity in the direction of the Firecracker.  Figure \ref{fig11} shows the spatial distributions of the bridge features {($V_{\rm LSR}$=1535--1560 km s$^{-1}$)} on the two velocity components in SGMC4--5 in a similar manner as in SGMC1. The bridges are distributed in the north and south of SSC B1. These components correspond to the two bridge features in velocity shown by the dashed lines in Figure \ref{fig8}. In summary, we find two velocity components and bridge features in {SGMC1, 2, and 4--5 associated with SSCs in the overlap region.}
{The detailed position-velocity diagrams of whole SGMC1, SGMC2, and SGMC4--5 are shown in Figures 20, 21, and 22, respectively.}

\section{{Application of displacements between the two interacting clouds}}

\begin{figure*}[htbp]
\begin{center}
\includegraphics[width=\linewidth]{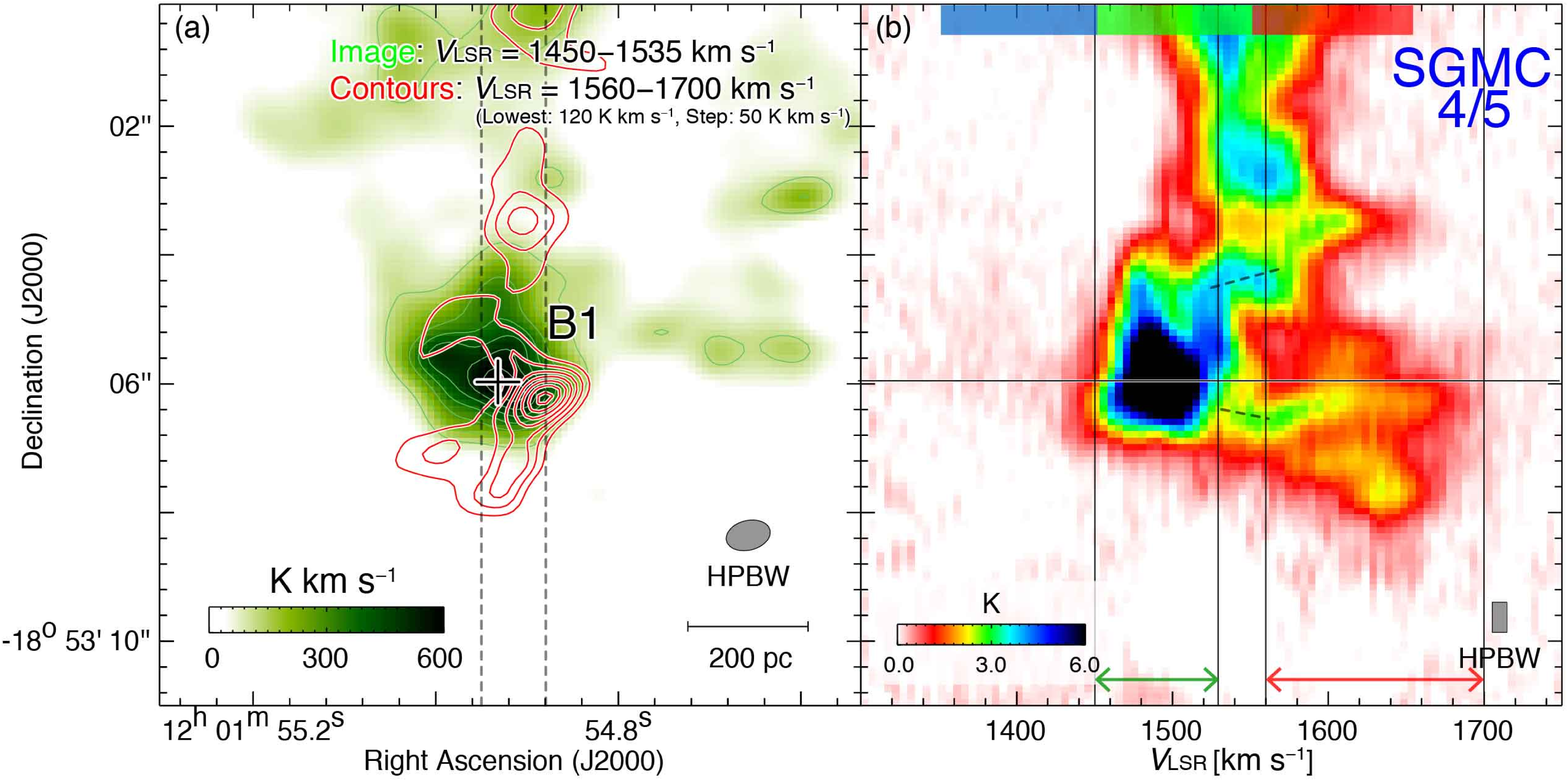}
\end{center}
\caption{(a) $^{12}$CO (3--2) intensity map of red-cloud (red contours) superposed on the green-cloud toward SGMC 4--5. The integration velocity range is $V$$_{\rm LSR}$ = 1580--1690 km s$^{-1}$ for the red-cloud and $V$$_{\rm LSR}$ = 1470--1530 km s$^{-1}$ for the green-cloud. The lowest contour level and intervals are 85 and 40 K km s$^{-1}$, respectively. The black cross indicates the position of SSC B1 (\cite{2007ApJ...668..168G}). The dashed lines show the integration range in R. A. in (b). (b) Declination-velocity diagram of $^{12}$CO (3--2).  The {black vertical lines} indicate the velocity ranges of the green- and red-clouds, respectively. The horizontal line indicates the declination of B1. The black dashed lines show the bridge features in the intermediate velocity range between the green- and red-clouds.} 
\label{fig8}
\end{figure*}

\begin{figure*}[htbp]
\begin{center}
\includegraphics[width=\linewidth]{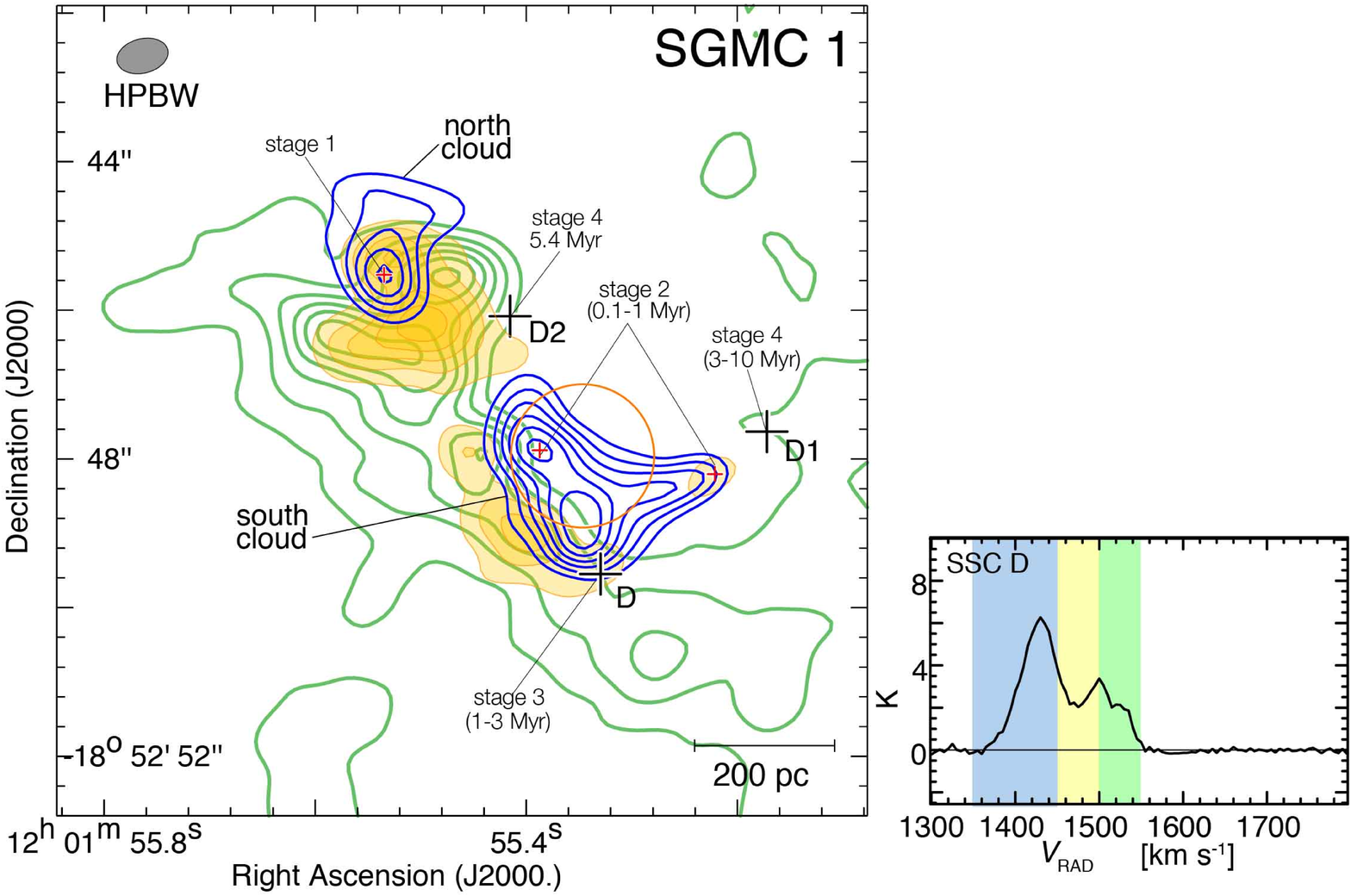}
\end{center}
\caption{$left$: Intensity map of $^{12}$CO (3--2) consisting of three velocity components of SGMC1. The {yellow} image indicates the intermediate velocity components (bridge; $V$$_{\rm LSR}$ = 1460--1505 km  s$^{-1}$). {The green and blue contours indicate the green-cloud ($V$$_{\rm LSR}$= 1505--1550 km  s$^{-1}$) and the blue-cloud ($V$$_{\rm LSR}$ = 1350--1460 km  s$^{-1}$), respectively. }The lowest contour level and intervals are 40 and 50 K km s$^{-1}$ for the green-cloud, 110 and 50 K km s$^{-1}$ for the intermediate velocity component, and 150 and 50 K km s$^{-1}$ for the blue-cloud, respectively. Black cross indicates the position of SSCs. {Red crosses indicate the regions where star cluster formation and evolutionary stages are classified by radio continuum, CO, and optical/NIR data (\cite{2014ApJ...795..156W}).} {$right$: the typical spectrum of $^{12}$CO (3--2) toward SSC~D. The blue-, yellow-, and green-ranges indicate the velocity ranges of the blue-cloud, bridge components, and the green-cloud, respectively.}} 
\label{fig9}
\end{figure*}

{The present results indicate that the clouds show complementary spatial distributions and bridge features in the Antennae, where a hint of displacements in the complementary distributions is also recognized (e.g., B1 in Figure \ref{fig8}a). As a next step we quantify a displacement in the present clouds, where the method was developed for quantifying a collision path in M43 (Fukui et al. 2018a). {The two features are typical signs of cloud-cloud collision as reviewed in Section 1.} The displacement is caused by the ballistic collision orbit which makes an angle to the line of sight significantly deviated both from 0 deg. and from 90 deg. and is proportional to the projected path traveled after a collision. We derive the displacements in the three regions in this Section and discuss the implications of the results in Section 5. }

\subsection{{SGMC1 and SGMC4-5}}
{In SGMC1 and SGMC4-5,} we calculated Spearman's correlation coefficient between the integrated intensity of the two complementary velocity components, and searched for a position where the correlation coefficient becomes minimum in order to derive an optimum displacement between the two complementary components as indicated by an arrow {in Figure12 and 13} (for details of the method see \cite{fujita2019binprep}). Specifically, we moved the red-cloud from the original position to $\pm$240 pc with a 1.8 pc (size of pixel) step, and calculated the correlation coefficient with the integrated intensity of the green-cloud for each pixel in such a way that the two components spatially coincide {with each other} after the displacement. As a result, we found the displacements in SGMC1 and SGMC4--5 {as shown in Figures 13 and 14.} 


In Figure \ref{fig12}b, the green-cloud shows an intensity depression toward the blue-cloud at the two positions as indicated by the black box in SGMC 1. We found that the two CO clouds exhibit a complementary distribution with displacements of 68 pc (P.A.=51.3 deg) and 92 pc (P.A.=54.5 deg.), respectively. These parameters give a minimum correlation coefficient of $-0.73$ and $-$0.81 {for the two regions, respectively}. 
In Figure \ref{fig13}b, the circular red-cloud and green-cloud are partly overlapped in SGMC 4--5. The green-cloud is surrounded by the red-cloud and they exhibit a complementary spatial distribution with a displacement of $\sim$63 pc in length (P.A.=$-$38 deg) as shown in Figure \ref{fig13}a {with a minimum correlation coefficient of $-$0.73}. 

\subsection{High resolution spatial- and velocity-distributions toward the proto-cluster cloud “{the} Firecracker”}
{In order to resolve the complementary spatial distribution toward {the} Firecracker, we analyzed }the spatial distribution of the CO cloud in SGMC2 by using the Cycle 4 data {(Finn et al. 2019)} whose resolution is $\sim$3 times higher than Cycle 0 data {in Figure \ref{fig7}. The Cycle 4 data in} Figure \ref{fig14}a {confirms the }complementary distribution of two velocity components {in Figure \ref{fig10}. In Figure \ref{fig14}a}  we redefined the two velocity ranges from [1400--1530 km s$^{-1}$ and 1580--1700 km s$^{-1}$] to [1370--1510 km s$^{-1}$ and 1595--1745 km s$^{-1}$] by using high-resolution 1st moment map. The red-cloud is located along the edge of the blue-shifted cloud{, showing a remarkable complementary spatial distribution at a 10-pc scale between the two clouds.}

 Figure 15b show{s} an enlarged view toward {{the} Firecracker}, where the two-velocity clouds in the same velocity range as above are shown. It is notable that the two distributions show significant difference with each other; the blue-shifted cloud is singly peaked and the red-shifted cloud has two peaks with an intensity depression between them. This suggests that the two clouds show complementary distribution even in a scale of 50 pc. {By applying the same method as above,} We find a $\sim$30 pc displacement {fits} the complementary distribution {as shown by an arrow.} Figure \ref{fig14}d shows a position-velocity diagram toward the proto-cluster cloud, and the two velocity components are connected by a bridge feature in velocity {at X$\sim$0.2 arcsec}. {The origins of the X and Y coordinates are set to be the position of the Firecracker, and the position angle of the Y axis is 55 deg from the north.} These results demonstrate that Firecracker shows the two signatures of cloud-cloud collision according to the method, {while the resolution is about 40\% of the displacement in Figure \ref{fig14}c. }

\section{Discussion}
\subsection{Collisions in the three complexes SGMC1, SGMC2, and SGMC4--5}

\begin{figure*}[htbp]
\begin{center}
\includegraphics[width=\linewidth]{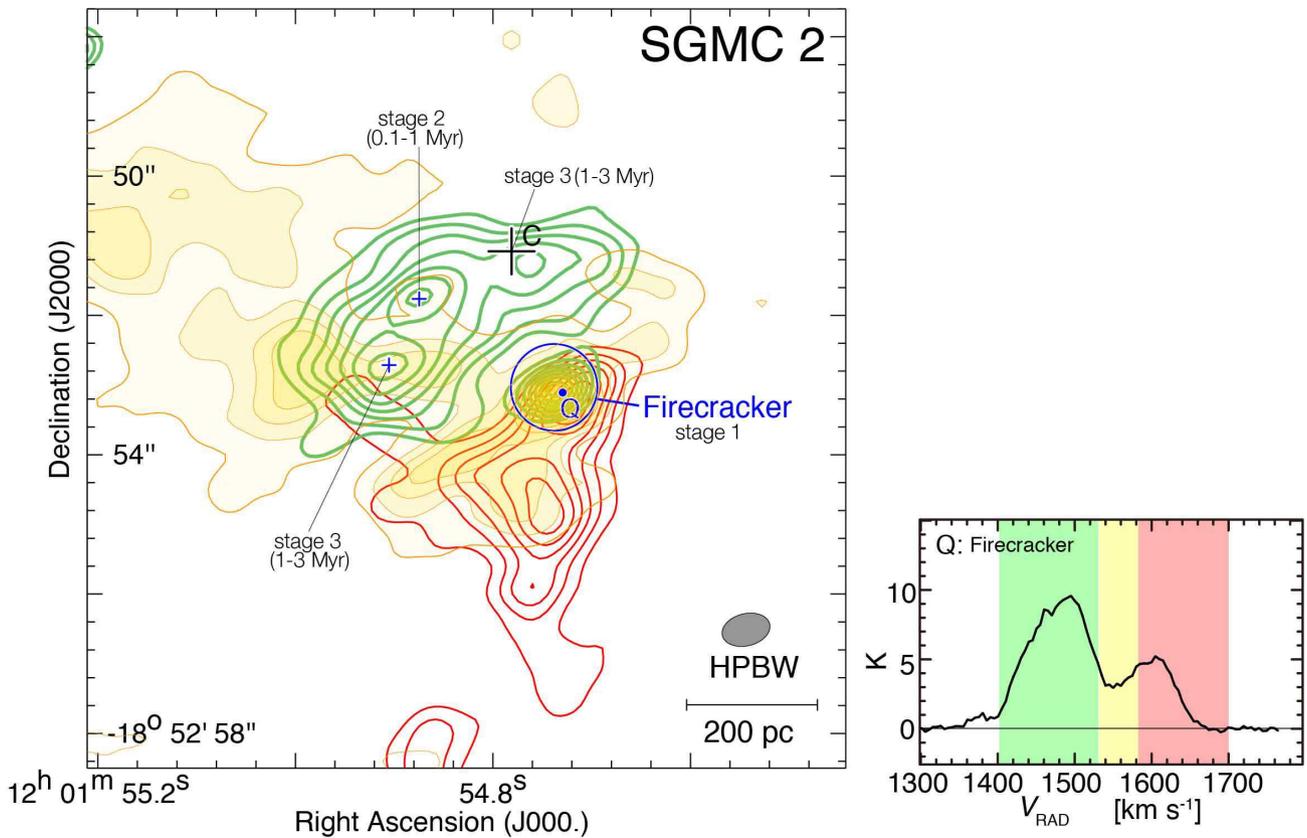}
\end{center}
\caption{$right$: Intensity map of $^{12}$CO (3--2) consisting of three velocity components around {the} Firecracker. The {yellow} image indicates intermediate velocity components (bridge; $V$$_{\rm LSR}$ = 1530--1580 km  s$^{-1}$). The green and red contours indicate the green-cloud ($V$$_{\rm LSR}$= 1580--1700 km  s$^{-1}$) and the red-cloud ($V$$_{\rm LSR}$ = 1580--1700 km  s$^{-1}$), respectively. The contour levels of red-cloud is the same as in Figure \ref{fig7}a. The lowest contour level and intervals are 25 K km s$^{-1}$ and 25 K km s$^{-1}$ ($\sim$10 $\sigma$) for the intermediate velocity component and 60 and 50 K km s$^{-1}$ for the green-cloud, respectively. Black cross indicates the position of SSC C. {Blue crosses indicate the regions where star cluster formation and evolutionary stages are classified by radio continuum, CO, and Optical/NIR data (\cite{2014ApJ...795..156W}).} {$right$: the typical spectrum of $^{12}$CO (3--2) toward {the} Firecracker. The green-, yellow-, and red-ranges indicate the velocity ranges of the green-cloud, bridge components, and the red-cloud, respectively.}} 
\label{fig10}
\end{figure*}

\begin{figure}[htbp]
\begin{center}
\includegraphics[width=\linewidth]{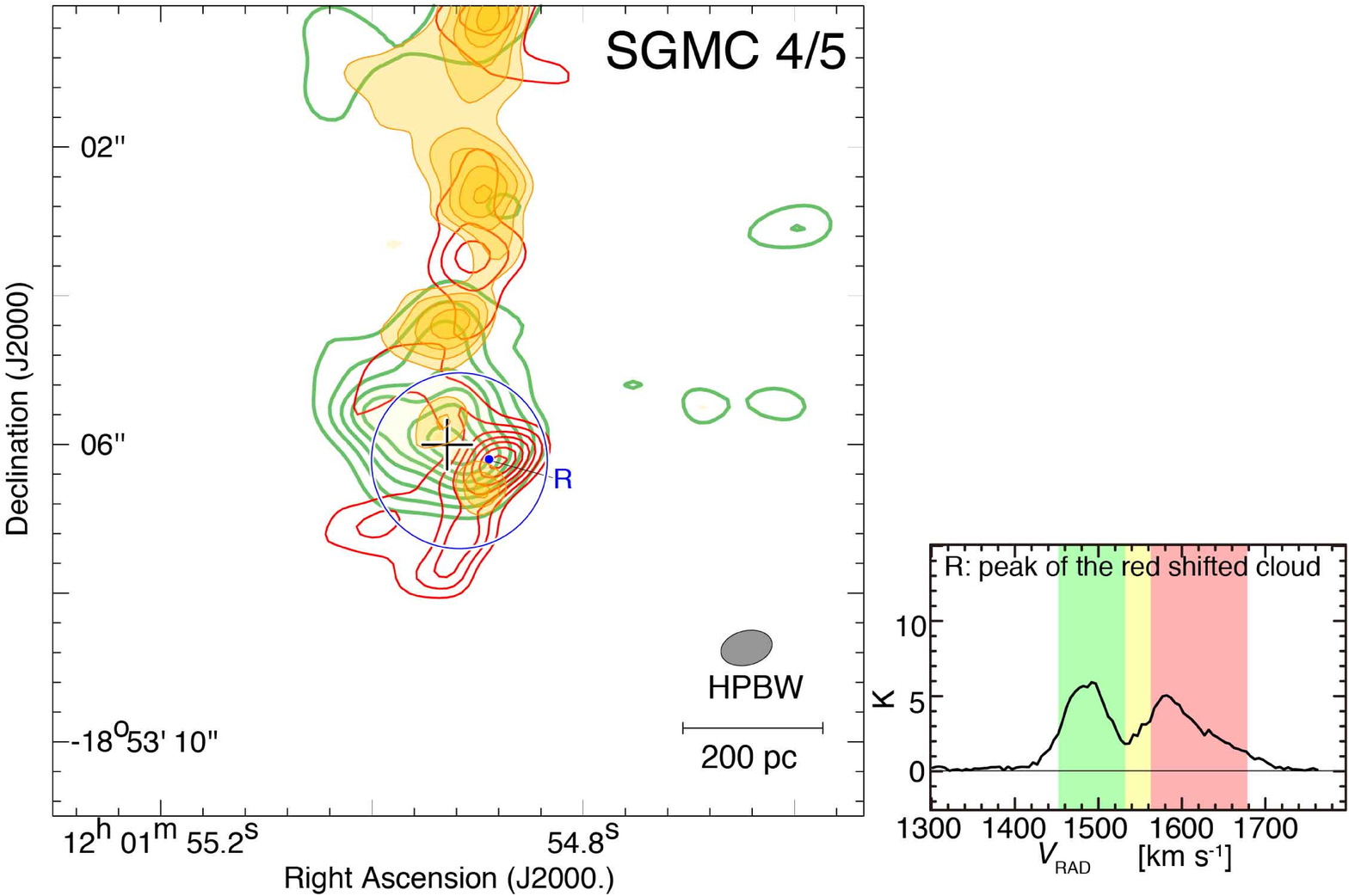}
\end{center}
\caption{$right$: Intensity map of $^{12}$CO (3--2) consisting of three velocity components around B1. The {yellow} image indicates intermediate velocity components (bridge; $V$$_{\rm LSR}$ = 1535--1560 km  s$^{-1}$). {The red and green contours indicate the red-cloud ($V$$_{\rm LSR}$= 1560--1700 km  s$^{-1}$) and the green cloud ($V$$_{\rm LSR}$ = 1530--1580 km  s$^{-1}$), respectively.} The contour levels of red-cloud is the same as in Figure \ref{fig8}a. The lowest contour level and intervals are 130 and 20 K km s$^{-1}$ for the intermediate velocity component and 85 and 40 K km s$^{-1}$ for the green-cloud, respectively. Black cross indicates the position of SSC B1. {$right$: the typical spectrum of $^{12}$CO (3--2) toward peak of the red-cloud. The green-, yellow-, and red-ranges indicate the velocity ranges of the green-cloud, bridge components, and the red-cloud, respectively.}} 
\label{fig11}
\end{figure}

The Antennae Galaxies is prominent because of the unusually active star formation including {thousands} of SSCs {with mass larger than 10$^5$ $M_{\rm \odot}$ (\cite{2010AJ....140...75W})}, and clear signatures of the galactic interaction including the two long tails. The ALMA CO data offer an ideal opportunity to investigate the molecular gas distribution and kinematics which are possibly related not only to the feedback but to the cluster formation. The previous works on the ALMA data focused mostly on the feedback by the clusters (\cite{2014ApJ...795..156W,2012A&A...538L...9H}), whereas the formation mechanism of the clusters was not pursued into depth. The CO clouds have a large velocity span of $\sim$200 km s$^{-1}$, which were interpreted as due to the cluster feedback by these authors. Except for a brief suggestion on a possible role of filamentary cloud collision in triggering cluster formation (\cite{2014ApJ...795..156W}), no exploration of the cluster formation was undertaken in these works. Subsequently, \citet{2015ApJ...806...35J} discussed the cluster formation mechanism in the Firecracker based on velocity integrated intensity of CO ALMA Cycle 0 data. \citet{2019ApJ...874..120F} presented ALMA Cycle {4} data on the Firecracker {and} {further discussed the cluster formation.} These authors tried to fit the radial profile of the column density by both a Gaussian and a Bonnor-Ebert spherical profile and derived the radial density distribution. They also discussed a possibility of cloud-cloud collision by referring to recent papers on cloud-cloud collision (e.g., \cite{2018ApJ...859..166F}). These previous works however did not attempt to apply a kinematic analysis which was undertaken in the present paper, and 
{did not make detailed modeling of the collision} in the cloud kinematics.

\begin{figure}[htbp]
\begin{center}
\includegraphics[width=6cm]{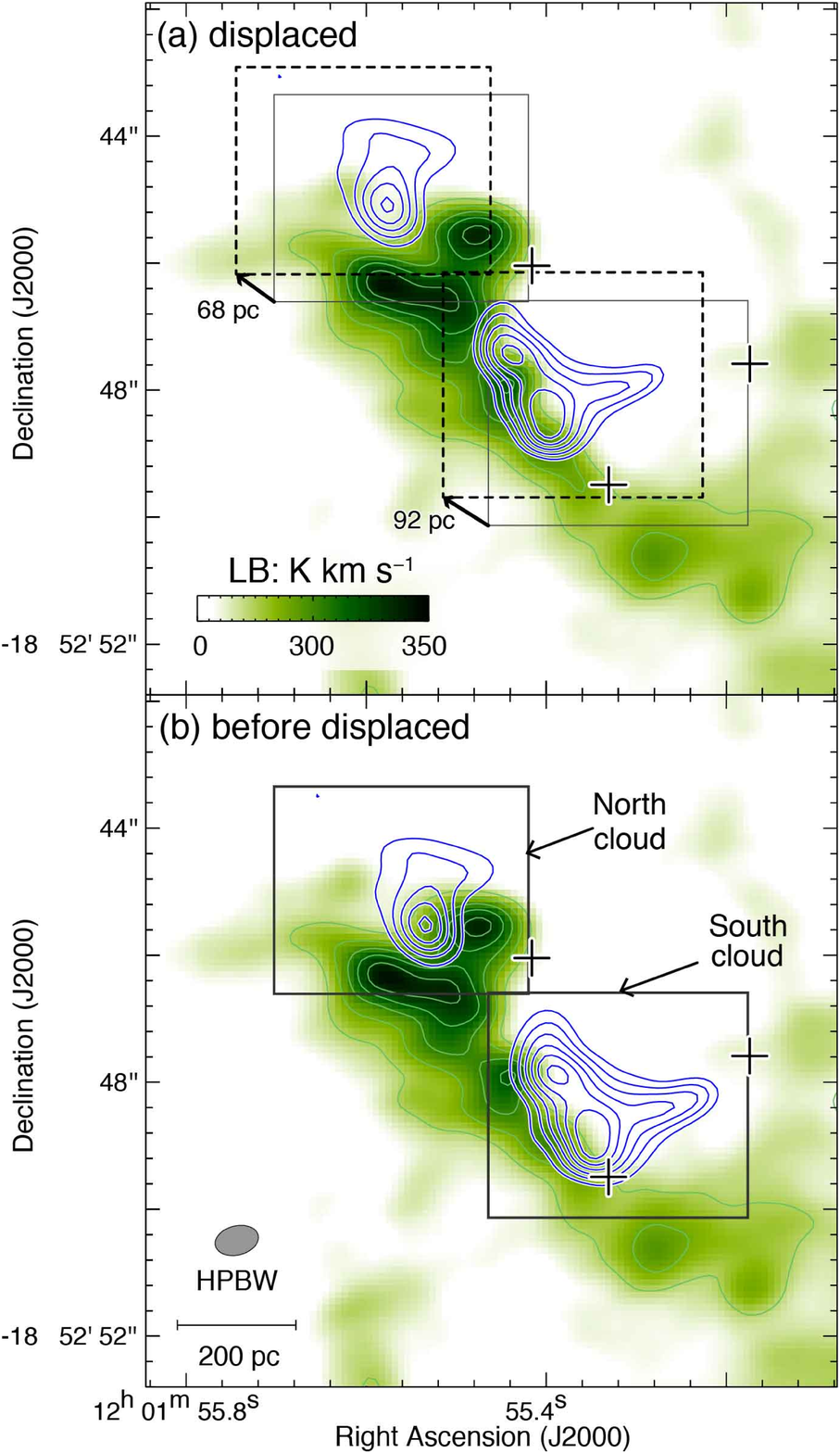}
\end{center}
\vspace*{-0.5cm}
\caption{Images of (a) and (b) are the same as in Figure {\ref{fig6}a}, whereas contours of the green-cloud in (a) are displaced. The projected displacement is 68 pc with a position angle of 51.3 deg. for the north cloud. The projected displacement is 92 pc with a position angle of 54.5 deg. for the south cloud. The black boxes denote the initial position of the green-cloud, and the dashed black box of (a) denotes the displaced position. The contour levels and symbol are the same as in Figure \ref{fig6}.} 
\label{fig12}
\end{figure}%

Recent studies in the Local Group galaxies suggest that the formation of the SSCs of 10$^4$--10$^5$ $M_{\odot}$ in the Milky Way, the LMC and M33 is likely triggered by a cloud-cloud collision; in the Milky Way cloud-cloud collisions are discovered in five SSCs Westerlund~2, NGC~3603, RCW~38, DBS[2003]179, and Trumpler~14 (\cite{2009ApJ...696L.115F,2014ApJ...780...36F,2016ApJ...820...26F,kuwahara2019inprep,fujita2019ainprep}) and in the LMC and M33 the remarkable cluster formation induced by the galactic tidal interaction is found in R136, N44 and NGC~604 (\cite{2017PASJ...69L...5F,2019ApJ...871...44T,2018PASJ...70S..52T}). Such triggering of high-mass cluster formation by cloud-cloud collision is in addition discovered in more than 50 regions of single or multiple high-mass star formation in the Milky Way {(see Table 2 of \cite{2019PASJ..tmp..127E})} by applying the new methodology to identify a cloud-cloud collision developed by \citet{2018ApJ...859..166F}. It is therefore important to test if a cloud-cloud collision between two molecular clouds is a viable scenario to trigger the cluster formation in the Antennae.

The present analysis shows that the three cloud complexes have observational signatures of two colliding clouds, i.e., the complementary distributions and the bridge features. As shown in Table 2, each of the complexes has molecular mass of (4--24)$\times$10$^7$ $M_{\odot}$ ($X_{\rm CO}$ mass) and a size of 300--500 pc. We present here a scenario that the two clouds in the three complexes are colliding with each other, and the collisional compression triggered the formation of the six SSCs/SSC candidate. The distributions of SGMC1 and SGMC4--5 show a displacement around 100 pc. A ratio of the displacement and the relative velocity gives $\sim$1 Myr (=100 pc/100 km s$^{-1}$) as the typical collision time scale over the complexes with no correction for the possible projection effect{, which is consistent with the ages of the SSCs B1, D, and the embedded clusters cataloged by Whitmore et al. 2014. Currently observed SGMCs are supposed to be the sites of the youngest cluster formation. However, the galactic interaction is considered to continue for more than several Myrs according to numerical simulations (e.g., Renaud et al. 2015). Hence it is suggested that the more evolved SSCs D1 and D2 were formed by similar past collisions.}

{Values of displacement and collision time scale are most affected by the colliding angle of two clouds $\theta$ relative to the line of sight{, where theta takes a value between 0 deg and 90 deg}. We can estimate the collision time scale $t$ following the equation as follows:}
\begin{eqnarray}
t=\left(\frac{A}{\rm sin \ \theta}\right)/\left(\frac{V}{\rm cos \ \theta}\right)=\frac{A}{V} \ \frac{1}{\rm tan \theta},
\end{eqnarray}
{where $A$ and $V$ are the projected displacement and the observed velocity separation between two clouds, respectively.} 
According to the numerical simulations {of an Antennae-like galaxy mergers by Renaud et al. (2015), the majority (80 \%) of the 
objects have high velocity dispersions larger than 100 km s$^{-1}$ reflecting the galactic-scale motion originated from the relative velocity of the two galaxies.}
Thus, we assume that the relative velocity difference between the two clouds is around {100--200 km s$^{-1}$, }{and the angle of the collision path to the line of sight $\theta$ should be smaller than 60 deg.}
{The angle of the collision axis $\theta$ with respect to the line of sight should be smaller than 60 deg, because for the case of vertical collisional motion ($\theta$=90 deg.), we do not see the two velocity components. For the case of collision along the line of sight ($\theta$=0 deg.), we do not see the displacement between the two velocity components. We thus can assume $\theta$ is 30--60 deg. The ideal case of $\theta$= 45 deg. collision gives a collision time scale of 0.6 Myrs for SGMC 4--5.}
Since the interaction velocity is not likely more than 200 km s$^{-1}$ for the galaxy mass of the Antennae, we roughly estimate $\theta$ is in a range from 30 deg. to 60 deg. corresponding to the time scale in a range of 0.3-1.2 Myrs. 
This gives an uncertainty of 0.6$^{+0.6}_{-0.3}$ Myrs. {We made similar estimations for the north- and south-clouds of SGMC~1 shown in Figure 13b. The time scale of north- and south-clouds are obtained to be 0.6$^{+0.6}_{-0.3}$ Myrs and 0.9$^{+0.9}_{-0.45}$ Myrs, respectively.}

\begin{figure*}[htbp]
\begin{center}
\includegraphics[width=14.5cm]{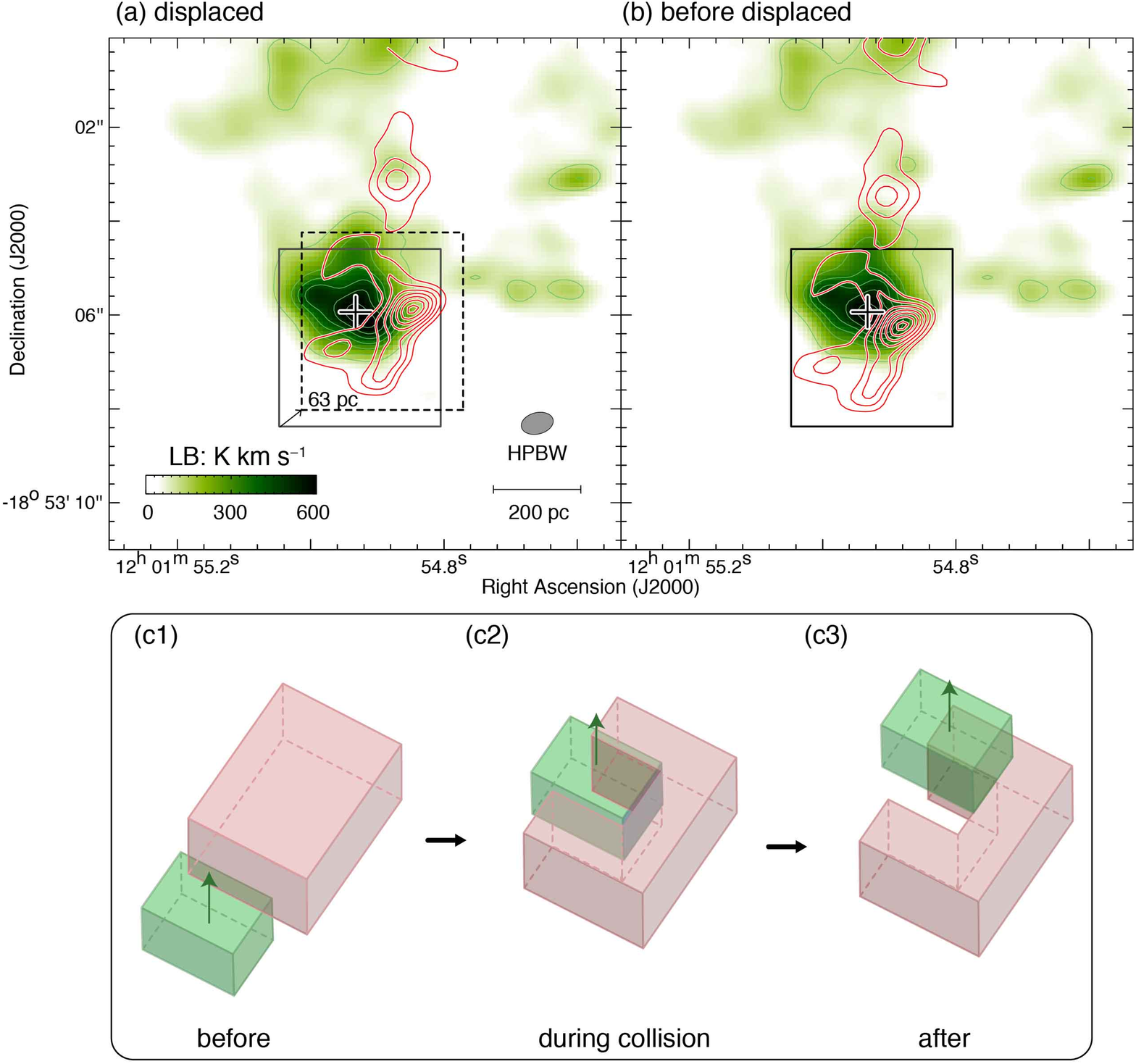}
\end{center}
\vspace*{-3.0cm}
\caption{Images of (a) and (b) are the same as in Figure \ref{fig8}a, whereas contours of the red-cloud in (a) are displaced. The projected displacement is 63 pc with a position angle of {$-$38 deg}. The black boxes denote the initial position of the red-cloud, and the blue box of (a) denotes the displaced position. The contour levels and symbol are the same as in Figure \ref{fig8}. Lower panel shows rectangular solid model clouds before the collision (c1), during the collision (c2), and after the collision (c3) for SGMC 4--5. {Panels} (a) and (b) correspond to (c2) and (c3), respectively.} 
\label{fig13}
\end{figure*}%

\subsection{Details of the collisional triggering in B1 and the Firecracker}
Among the SSCs, B1 is the most luminous embedded cluster candidate in the Antennae, and the mass of the cluster is estimated to be $\sim$10$^6$ $M_{\odot}$ from the high H$_{2}$ luminosity 10$^8$ $L_{\odot}$ (\cite{2012A&A...538L...9H}). The cloud mass in {SGMC4--5} is estimated to be $\sim$10$^7$ $M_{\odot}$ with a size of $\sim$200 pc at a 20\% level of the peak integrated intensity (Table 2). The Firecracker has no luminous source yet and it is suggested that the gas cloud is in a proto-cluster stage and {might} form a SSC soon (\cite{2014ApJ...795..156W}).

A possible scenario for the complementary distribution of the SSC B1 in SGMC4--5 (Figures 14a and 14b) is that the {green-}cloud collided with the central part of the {red-}cloud and created the cavity in the red-cloud. A schematic is shown as three collision steps, prior to, during and after the collision, in Figure \ref{fig13}. {Moreover, the collision time scale of 0.3--1.2 Myr is roughly consistent with the age of SSC B1 1--3.5 Myr as shown in Table 3.}

{The morphological appearances of the clouds experienced collisions may greatly depend on the initial conditions of clouds such as density distributions and collision velocity. When a low-density cloud collides with a denser one, the majority of the low-density gas is expected to be dispersed as a result of momentum conservation. As shown in Section 5.3, the collision of SGMC2 is off-centered and only the northern part of the red-cloud experienced collision in SGMC2, whereas that of SGMC4--5 is head-on and 
the entire red-cloud in SGMC 4--5 experienced the collision. The velocity structures shown in Figure 9b imply that the initial density of the red-cloud was low and it has been mostly dispersed by the head-on collision with the dense blue-cloud.} {In addition, we test here the dynamical binding of the green- and red-clouds of SGMC4--5. The total mass required to gravitationally bind these two clouds can be calculated as $M$=$l$ $\sigma$$^2$/2G$\sim$3$\times$10$^8$ $M_{\rm \odot}$, where $l$ is separation of the two clouds, and $\sigma$ is velocity separation of the two clouds. We assume that the two clouds are separated by 89 pc in space as shown in Figure 14 and by 156 km s$^{-1}$ in velocity (assuming the collision angle to the line of sight $\theta$ is 45 deg). This value is a factor of three larger than the total molecular mass of 8.5$\times$10$^7$ $M_{\rm \odot}$ calculated in section 3.2. This indicates that the coexistence of the two velocity clouds are not in a gravitationally bound system.}

\begin{figure*}[htbp]
\begin{center}
\includegraphics[width=16.5cm]{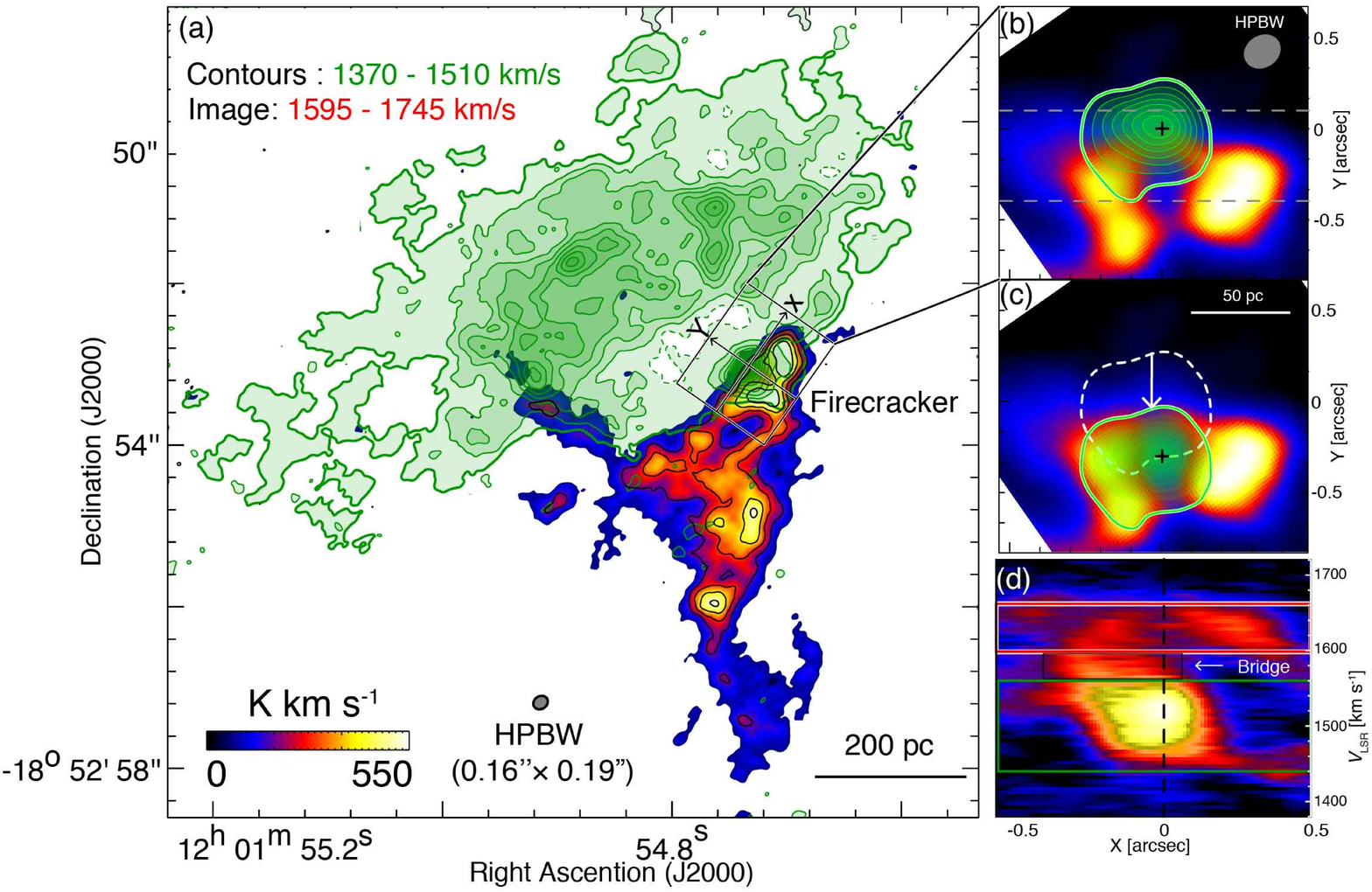}
\end{center}
\vspace*{-0.5cm}
\caption{(a) $^{12}$CO(3--2) intensity map of two colliding clouds toward SGMC2 shown in grey box of Figure 4 obtained with ALMA Cycle 4 project (\#2016.1.00924.S, \cite{2019ApJ...874..120F}). The average angular resolution is 0.13 arcsec ($\sim$14 pc). The contours and image show the blue shifted cloud and red shifted cloud, respectively. The lowest contour level and interval are 5$\sigma$ and 100 K km s$^{-1}$, respectively. (b) Enlarged view toward {the} Firecracker. The contours and image show the blue shifted cloud and red shifted cloud, respectively.  {The integrated velocity ranges of two clouds are justified to be 1370--1510 km s$^{-1}$ and 1595--1745 km s$^{-1}$ by using high spatial resolution moment map.} The black cross indicates the position of the Firecracker (R.A.={$\timeform{12h1m54.7344s}$ , Dec.=$\timeform{-18D52'53.0184"}$}). The position of {the} Firecracker is set as the origin of XY coordinate, and the position angle of the Y axis is 55 deg. (c) Same as (b), but the contour of the blue shifted cloud is displaced {by} $\sim$ 30 pc. (d) P--V diagram toward {the} Firecracker. The black perpendicular line indicates the position of {the} Firecracker. The grey dashed lines in (b) show the integration range of P--V diagram. {The red- and green-boxes show integrated ranges of the red- and green-clouds, respectively.}} 
\label{fig14}
\end{figure*}

A similar scenario is applicable to the Firecracker. Figures \ref{fig14}b and \ref{fig14}c show the distribution of the two clouds {toward {the} Firecracker} produced from the ALMA Cycle 4 data by \citet{2019ApJ...874..120F}. The morphology of the collision in the Firecracker is very similar to B1 (Figure \ref{fig13}c), while the size of the colliding clouds is about a factor of three smaller than in B1, and the mass of the small cloud, i.e., the intermediate velocity (IM) cloud {between the green- and red-clouds}, is about 10 times smaller than in B1. {The mass of the IM cloud toward B1 and the Firecracker are $\sim$3.5$\times$10$^7$ $M_{\odot}$ and $\sim$3.5$\times$10$^6$ $M_{\rm \odot}$, respectively.} In the two objects, the bridges connecting the two clouds are shown in Figures 8b, 9b, 11, 12, and \ref{fig14}d. It is possible that after the collision the mass in the cavity and the blue-shifted cloud is converted into part of the IM cloud and the bridge as well as the cluster member stars. More details of the mass budget depend on the initial cloud distribution {(\cite{2014ApJ...792...63T})} and is currently uncertain at best.


A ratio between the displacement and the velocity separation indicates a collision time scale of {0.15--0.6} Myr with uncertainty of a factor of {two} due to the projection effect, which is consistent with {the time scale of the collapse of the molecular cloud and {no star formation} of {the} Firecracker (\cite{2015ApJ...806...35J})}. 
In {our proposed} scenario, the IM cloud is highly turbulent as a result of the momentum injection in the collision at 100 km s$^{-1}$. According to the numerical simulations of colliding molecular flows (cf. \cite{2013ApJ...774L..31I}), the velocity dispersion is produced by the momentum injection between the colliding clouds and the velocity dispersion is close to their velocity separation. The turbulence then causes high pressure in the parent cloud forming the cluster, which increases the Jeans mass to a high value. The pressure inferred in the IM cloud is estimated to be $\sim$10$^8$/$k_{B}$ {K cm$^{-3}$} for velocity dispersion of 30 km s$^{-1}$ and average molecular density of 10$^2$ cm$^{-3}$ as estimated by \citet{2019ApJ...874..120F}, in accord with the formation criterion for a cluster of 10$^6$ $M_{\odot}$ (\cite{1997ApJ...480..235E}). Further observational details of the cluster formation must await higher resolution data with ALMA. In the Firecracker, the molecular mass of the small cloud is 2$\times$10$^7$ $M_{\odot}$ within $\sim$37 pc {of radius (Table 2)}, several times smaller than that that of B1, suggesting that the cluster possibly forming in the Firecracker {will have a} mass smaller than B1. The collision time scale is here estimated to be 0.3 Myr (= 30 pc/110 km s$^{-1}$), shorter than in B1, being consistent with the proto-cluster nature if the projection effect is similar between them. The schematic in Figure \ref{fig13}c indicates that the cluster forming region is probably on the near side of the small cloud which strongly interacted with the other cloud. {We may be able to test the schematic of SGMC 4--5 by comparing the higher resolution ALMA data with $HST$ image.}

\begin{figure*}[htbp]
\begin{center}
\includegraphics[width=\linewidth]{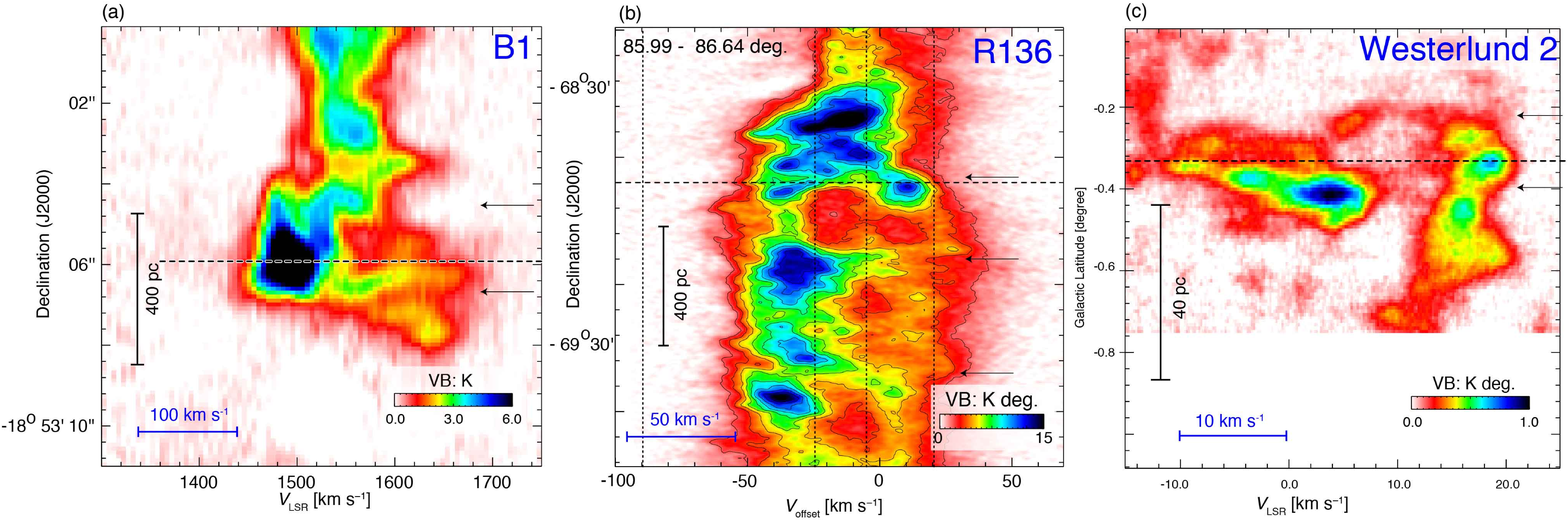}
\end{center}
\caption{Position-velocity diagrams for SSCs in the Antennae Galaxies, the Large Magellanic Cloud, and the MW. (a) Declination-velocity diagram of $^{12}$CO (3--2) for B1. (b) Declination-velocity diagram of H{\sc i} for R136 obtained with Australia Telescope Compact Array \&Parkes telescope (Kim et al. 2003). (c) Declination-velocity diagram of $^{12}$CO (2--1) for Westerlund 2 (Furukawa et al. 2009; Ohama et al. 2010).   The black arrows in (a), (b), and (c) denote the position of bridge features. {The horizontal dashed lines indicate the positions of SSCs.}} 
\label{fig15}
\end{figure*}%

\subsection{{Formation of SSCs in SGMC1 and SGMC 2}}
{In SGMC 1 and SGMC 2, the SSCs are located a little outside of their corresponding SGMC.} {The relation among the cloud-cloud collision, the cluster formation, and its evolutionary stage toward SGMC~1 and SGMC~2 is discussed in this Section. Whitmore et al (2014) defined the five evolutionary classifications of the SGMCs. Based on the previous study of the GMC evolution model in the Local Group of galaxies (LMC; Fukui et al. 1999, Kawamura et al. 2009, M33; Miura et al. 2012), the authors extended the classification to older clusters with no CO by using radio continuum, optical/IR data in addition to CO as follws: Stage 0 (diffuse giant molecular clouds), Stage 1 (centrally concentrated molecular clouds), Stage 2 (embedded cluster, 0.1--1Myr), Stage 3 (emerging cluster, 1--3Myr), Stage 4 (young cluster, 3 Myr--10 Myr), and Stage 5 (intermediate/old clusters, $>$10 Myr).} {We summarized physical properties of collision and evolutionary classification of the SGMCs in Table 3}.

{In SGMC~1, complementary spatial distributions and bridge features are observed, which are observational signatures of cloud-cloud collision as shown in Figure 6} { and Figure 7. Since SSC D2 is located at the edge of the green cloud, {It is plausible that the natal cloud has already been ionized and dispersed by the cluster, because its age is as old as 5.4 Myr (\cite{2007ApJ...668..168G}){, which is} derived from the observed equivalent width of Br$\gamma$ using the Starburst99 fittings. Thus, D2 may be related to the past collision between the blue- and green-clouds located 
in the northern part of SGMC~1.}
{There are bridge features between the green- and blue-clouds shown in Figure 7a, and it can be seen from Figure 10 that these bridge features spatially connect the blue- and green-clouds.} { Toward D2, the bridge feature is distributed so as to connect the peak of the blue-cloud and intensity depression of the green-cloud. {Thus, D2 may be related to the past collision between blue- and green-clouds located in the northern part of SGMC~1.} Toward {the} peak of the north cloud, there are observational signatures of collision and strong bridge features, but there {is} no cluster formation and classified as stage~1 by Whitmore et al (2014). {This suggests that next generation star formation will be triggered in this area, since the collision time scale is as short as 0.3--1.2 Myr.}

 {SSC D is located at the boundary between the blue- and green-clouds, and the bridge feature is also distributed toward D. }{There are bridge features at the right ascension where the SSCs exist as shown in vertical lines of Figure 7.} {The typical time scale is $\sim$1 Myr, which was derived by a ratio of displacement and the relative velocity of colliding clouds. It is consistent with the age of D (1.5 Myr; \cite{2010AJ....140...75W})} {and {its} evolutionary stage 3 (1--3 Myr).}

 As for SSC D1, the blue- and green-clouds are connected by bridge features in a velocity space toward D1 shown by black dashed lines in Figure 7b, {but there is little evidence of natal molecular cloud,} since {its} evolutionary stage is 4 (3--10 Myr). 
 There are two embedded clusters according to Whitmore et al (2014) {in the} east of D1 {as} shown by red crosses in Figure 10. These two regions are observed by radio continuum and optical/IR emissions, and classified as embedded cluster (stage 2: 0.1--1 Myr). {The blue- and green-clouds are connected by the bridge features toward the early-stage potential cluster as shown in Figure 10.} Since there are collision signatures shown in Figure{s} 6 and 7b toward these regions, cluster formation due to collision may be triggered.

As for the whole SGMC1, {cluster formation is only apparent on the western side of the green cloud, implying that the collision occurred only on the west{ern} side, {but} not {on} the east{ern side}.}
In Section 5.2, we focused on the collision signatures at 50 pc scale in the direction of {the} Firecracker, but collision also occurred at 20--500 pc scale toward SGMC~2. As you {is} see{n in} from Figures 8{--}\ref{fig14}, {the} edge of the red-cloud is well fitted to the boundary of {the} green cloud. This signature can be interpreted as an off-center collision rather than the head-on collision {as} calculated by Takahira et al. (2014). The green-cloud (small cloud) does not form a cavity in the red-cloud (large cloud) like B1 and {the} Firecracker as shown in Figures 14 and \ref{fig14}, but it is interpreted as collision between the green-cloud and the northern {edge} of {the} red-cloud. This off-center collision {may be} similar to the results of {a} cloud-cloud collision in the Galactic H{\sc ii} region of NGC 2068/2071 (Fujita et al. 2019){, where new simulations} of an off-center collision {are shown}. The spatial and velocity structures of SGMC~2 are similar to the results of numerical simulation of the off-center collision so it is {possible} that a collision similar to NGC2068/2071 occurred.

{SSC C is located 100--200 pc away from the boundary of {the} colliding two clouds. In the velocity space, there is a diffuse bridge feature at the declination of C (dashed horizontal line in Figure \ref{fig7}b), but the red-cloud and bridge features are not spatially distributed toward C as shown in Figure \ref{fig10}. This {may} be due to the dissipation of {the} molecular cloud because of the age of C is as old as 5--6 Myr. Furthermore, it is likely  that the signatures of collision {could be} observed {with} H{\sc i} even if the observational trends of collision are not observed by CO. }
{In fact, it has been found that the formation of molecular clouds and star clusters was triggered by tidally driven colliding H{\sc i} gas in the LMC and M33 (Fukui et al. 2017,18; Tsuge et al. 2019; Tokuda et al. 2018; Tachihara et al. 2018). Future observations {of H{\sc i} by next-generation telescopes} will enable us to elucidate detailed spatial and velocity structures of H{\sc i} gas. } 
{There {are} two more cluster forming regions on the eastern side of the green-cloud shown by blue crosses in Figure 11. There are bridge components in this area, and the green{-} and red-clouds are connected by bridge features in velocity space {as} shown by black dashed lines in Figure 21f. From these results, cluster formation is possibly triggered by collision in these two regions.} According to these observational signatures, {it is possible} that the cluster formation was caused by collision in both SGMC~1 and SGMC~2.

\subsection{{Possible role of stellar feedback in accelerating the gas motion}}

{It is necessary to consider the effect of stellar feedback on the motion and disruption of the cloud, because the star formation rate of the region toward SGMC1, SGMC2, and SGMC 45 is 1.08--2.78 $M_{\rm \odot}$ yr$^{-1}$ (\cite{2010A&A...518L..44K}), which is most active in the Antennae Galaxies. In this Section, we argue that the velocity difference between the two molecular clouds cannot be explained only by stellar feedback based on the observational parameters. \citet{2017A&A...600A.139H} proposed a possible scenario that the spatial distribution and velocity separation in SGMC4--5 (Figure \ref{fig8}) originated by radiation pressure as an outflowing gas with a velocity separation of $\sim$100 km s$^{-1}$. {This scenario is, however, unlikely} because the radiation pressure in the early stage is assumed to be several ten{s of} that of the present day based on model (\cite{2010ApJ...709..191M}). If we assume that the expansion velocity is about a half of the velocity separation, the expansion velocity is estimated to be $\sim$50 km s$^{-1}$. As for SGMC~4--5, the mass of the green-cloud is 6$\times$10$^7$ $M_{\odot}$ and that of the red-cloud is 4$\times$10$^7$ $M_{\odot}$, {based on a} CO $J$=3--2 /$J$=1--0 ratio of} {0.5$\pm$0.1 (\cite{2012ApJ...745...65U})} {and $X_{\rm CO}$ $\sim$0.6$\times$10$^{20}$ cm$^{-2}$ (K km s$^{-1}$)$^{-1}$ (\cite{2003ApJ...588..243Z,2014ApJ...795..174K}). For the total cloud mass $\sim$10$^{8}$ $M_{\odot}$, the total kinetic energy required {for the gas motion} is estimated to be 1.3$\times$10$^{54}$ erg. If we adopt an efficiency of conversion of a supernova explosion (SNe) into the gas kinetic motion to be 5\% (\cite{2012MNRAS.426.3008K}), the observed kinetic energy requires 26000 SNe's. {The} total stellar mass of B1 is $\sim$280 times that of Westerlund~2 whose stellar wind energy $\sim$3.6$\times$10$^{51}$ erg (e.g., \cite{2007A&A...463..981R}), so the expected stellar feedback is $\sim$10$^{54}$ erg. Assuming ideal adiabatic wind bubbles, 20\% of the wind luminosity is transferred to the expanding gas shell (Weaver et al. 1977). {Then, 2$\times$10$^{53}$ erg of stellar wind energy is converted into the expansion energy}, which is an order of magnitude smaller to explain the gas motion. In addition, the solid angle subtended by the clouds is small{er than 4$\pi$}, {reducing} the energy available to {the expansion}. }{Moreover, the cloud distributions sometimes show {a} circular shape, which is not spherical-shell like but is ring-like, showing no sign of expansion (e.g., \cite{2010ApJ...709..791B})}{, feeding only a fraction of the kinetic energy into the expansion. In summary, the stellar feedback energy is unlikely to cause expansion of the clouds even if the energy is totally converted to expansion.}

{We made similar calculations for SGMC~1 and SGMC~2 by using the cloud parameters in Table 2, and obtained total kinematic energies to be $\sim$1.4$\times$10$^{54}$ erg, and $\sim$10$^{54}$ erg considering the {efficiency} of conversion {of} luminosity to kinematic energy, respectively. These energies are more than that in SGMC~4--5 and would require even more supernovae by the same assumption above. These numbers of SNe seem to be too large in a time scale of 1Myr, unfavorable to the feedback interpretation. }

\begin{figure}[htbp]
\begin{center}
\includegraphics[width=8cm]{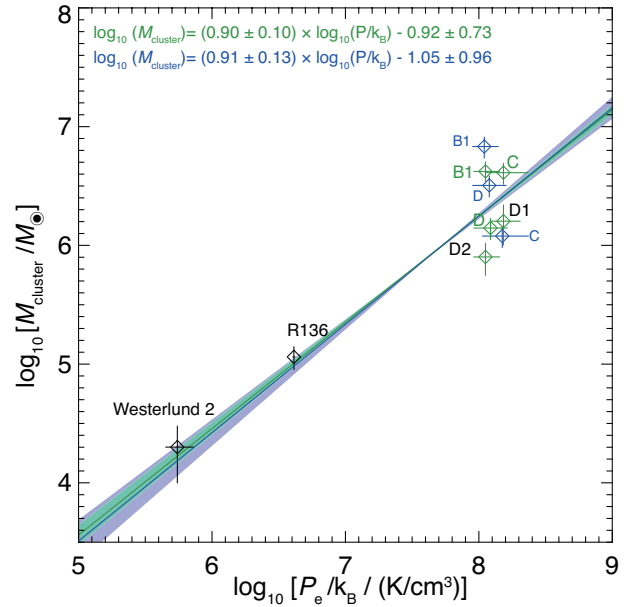}
\end{center}
\caption{{Correlation plot between the external pressure ($P_{\rm e}/k_{\rm B}$) and total stellar mass of SSC ($M_{\rm cluster}$). {Masses of SSCs in the  Antennae cataloged by \citet{2007ApJ...668..168G} are illustrated by the green diamonds. Blue diamonds indicate the masses of SSC B, C, and D cataloged by Whitmore et al. (2010). Representative values of the $P_{\rm e}/k_{\rm B}$ and $M_{\rm cluster}$ are plotted by the circles with error bars. The areas colored and the solid lines show variation of the correlation and the linear regression by $\chi^{2}$ fittings, respectively. Light green and blue indicate the results using the values of cluster mass cataloged by \citet{2007ApJ...668..168G} and Whitmore et al. (2010), respectively.}}} 
\label{fig16}
\end{figure}%



\subsection{Comparison with the other clusters }
We present in Table 3, Figures \ref{fig15}, and Figure \ref{fig16} {a} comparison of the physical parameters of the three regions where cloud-cloud collisions are suggested. They include B1, {C, D,}{D1, D2,} R136, and Westerlund~2, which show two velocity components at {16--120} km s$^{-1}$ velocity separation linked by bridge features{, respectively as shown in Table 3}. {It is suggested that high-pressure gas is required for the formation of globular clusters by both the observational and theoretical studies (\cite{1990A&A...234..156V,1997ApJ...480..235E}). } {This kind of high pressure environment is likely formed by interstellar shocking in the interacting galaxies. {We then} compared the external pressure and the total stellar mass of SSCs.}

{We calculated the external pressure at the cloud edge ($P_{\rm e}$) from the cloud's mass, size, and colliding velocity following the equation (\cite{1989ApJ...338..178E}) as follows:   
\begin{eqnarray}
P_{\rm e}= \frac{3\Pi \it M v^{2}}{4 \pi R^{3}} = \rho_{e} v^{2},
\end{eqnarray}
where $M$ is the cloud mass, $v$ is the colliding velocity (difference of peak velocity of colliding clouds), $R$ is the radius of cloud, and $\rho_{e}$ is the density at the cloud edge. $\Pi$ is defined by $\rho_{e}$=$\Pi$$\rho$, where $\rho$ is the mean density in the cloud. We adopt $\Pi$ = 0.5 (\cite{2015ApJ...806...35J,2019ApJ...874..120F}). We calculated the pressure by using the physical parameters summarized in Table 2 for }{B1 and {the} Firecracker. The pressure{s} of C, D, D1, and D2 are calculated from the region of 150 pc radius (roughly corresponding to the size of SGMC4--5) from the cluster since SGMC1 and SGMC2 are spread about 500 pc and contain many clouds that are not directly related to the formation of SSC.} {The error of the external pressure is dominated by the uncertainty of the mass estimation. For the cloud mass estimation, we assume the CO 3-2/1-0 ratio as which seems to have about 20--30 \% errors as shown in section 3.2. This gives the relative errors for the cloud density of 20--50\% with the average of $\sim$30\% .
The error of the external pressure is estimated from the uncertainties of $J$=3--2 /$J$=1--0 ratio, signal to noise ratio of H{\sc i}, and 1 $\sigma$ error of gaussian fitting. The uncertainties of 3--2/1--0 ratio are dominant in the estimated error. The relative errors of ratio are about 20--30 \% as shown in section 3.2. From these uncertainties, the relative errors of density of colliding clouds are estimated to be 20--50\%, and average value is $\sim$30 \%.}

{As for B1 and {the} Firecracker, $P_{\rm e}$ is higher in the green-cloud where the cluster exists than the red-cloud. This result is interpreted as due to the difference of density of colliding clouds. In the Firecracker, density of the green cloud is $\sim$5 times that of the red-cloud. {In} B1, {the} density of the green cloud is $\sim$3 times that of the red-cloud as shown in Table 3. Thus, it is suggested that the external pressure $P_{\rm e}$ of {the} denser {green-cloud} is enhanced in the compressed layer and induced the formation of SSCs.} {In the case of the Firecracker, the denser green-cloud induced the formation of SSC similar to B1. The formation of SSCs D, D1, and C are possibly associated with the green clouds with higher $P_{\rm e}$
The formation of SSCs D, D1, and C are possibly associated with the green clouds which have higher .  However it is difficult to judge which velocity component the formed clusters belong, because the compressed gas by the collision should have been decelerated and has intermediate velocity between the two initial velocities. }

{For R136, we calculated H{\sc i} masses of the L- and D-components {(\cite{2017PASJ...69L...5F,2019ApJ...871...44T})} at the position of (R.A., Dec.)=($\timeform{D84.506234}$--$\timeform{D85.025770}$, $\timeform{D-68.982965}$--$\timeform{D-69.103751}$) as 2.2 $\times$ 10$^{4}$ $M_{\odot}$ and 2.3 $\times$ 10$^{5}$ $M_{\odot}$, respectively. External pressures would be 4.1$\times$10$^6$ K cm$^{-3}$ for the L-component (radius of 25 pc) and 2.0$\times$10$^6$ K cm$^{-3}$ for the D-component (radius of 70 pc). We use the equation as follows (\cite{1990ARA&A..28..215D}): 
\begin{eqnarray}
N_{\rm H{\sc I}} = 1.8224\times10^{18}\int \Delta T_{\rm b}  \it dv \ \rm{[cm^{-2}] },
\end{eqnarray}
where $T_{\rm b}$ is the observed H{\sc i} brightness temperature [K]. }
{In the LMC, it is suggested that $\sim$400 O/WR stars in the south east of the LMC including R136, N159 and the other active star forming GMCs are triggered by the collision of H{\sc i} flows. Thus, it is suitable to calculate the pressure of H{\sc i} gas for R136.}

{The external pressure of Westerlund~2 is calculated by the physical parameters of molecular clouds (\cite{2009ApJ...696L.115F}). Colliding velocity $v$=16 km s$^{-1}$, and the masses of red cloud ($V_{\rm LSR}$=11--21 km s$^{-1}$) and blue cloud ($V_{\rm LSR}$=1--9km s$^{-1}$) are $\sim$1.0$\times$10$^{5}$ $M_{\odot}$ and $\sim$8.0$\times$10$^{4}$ $M_{\odot}$, respectively (Furukawa et al. 2009). So, the pressure would be $P_{\rm e}/k_{\rm B}$ = 5$\times$10$^{5}$ K cm$^{-3}$ for radius of $\sim$30 pc as shown in Table 3.}


{We made a correlation plot between the external pressure and total stellar mass of {these} SSCs as shown in Figure \ref{fig16}.  We plotted the maximum value of the external pressure for each SGMC since the cloud under higher external pressure is considered to be related to star formation. {The} Firecracker is not included in the correlation plot because it is a very young proto cluster cloud and at a different evolutionary stage{, and it has no measurable stellar mass, so could not make a direct comparison with the others.} We performed linear fittings using the MPFITEXY routine, which provides the slope and intercept values (\cite{2010MNRAS.409.1330W}). We finally obtain that the slope is 0.90 and the intercept is $-$0.92, and linear Pearson correlation coefficient is 0.96 {when we adopted masses cataloged by Gilbert \& Graham (2007). If we adopted masses cataloged by Whitmore et al. (2010), the results are almost unchanged as the slope is 0.91, the intercept is $-$1.04, and linear Pearson correlation coefficient is 0.93.} There is significant positive correlation between them.} {Typical relative errors in the total stellar masses are 20\%, which are dominated by the photometric errors for most of the sources. On the other hand, relative errors of D1 and D2 are 37\% and 30\%, respectively, due to their large age uncertainties.}

B1 has the largest cluster mass and the formation {may be} triggered by colliding CO flows at 100 km s$^{-1}$. Westerlund~2 also {may be} triggered by a collision of two CO clouds at ten times smaller velocity. The mass ratio of Westerlund~2 to B1 is $\sim$100. The pressure of the formation environment in B1 is {60--180} times higher than that in Westerlund~2. On the other hand, the pressure in R136 lies between the two which reflects the lower density of the H{\sc i} flows at 70 km s$^{-1}$ collision speed. {This suggests that a SSC can be formed {if} the pressure is enhanced enough even if the colliding clouds are low density.}
So, the sequence in cluster mass is consistent with the theoretical idea for cluster formation in the literature (\cite{1997ApJ...480..235E}). Collisional triggering at supersonic velocity is a natural and consistent scenario in the observational results obtained so far. 

{In order to elucidate the cloud formation, we need to have a deeper understanding of the physical process into more detail}; at present the MHD simulations by \citet{2013ApJ...774L..31I} of colliding molecular flows is limited to a low velocity less than 30 km s$^{-1}$. Future efforts {will} be made for higher velocity flows at 100 km s$^{-1}$, and include H{\sc i} collisions for our understanding of cluster formation over a full mass range.

\section{Conclusions}

The Antennae galaxies is the most outstanding major merger closest to the Milky Way. It shows extremely active star formation including {thousands} of young massive clusters, and clear signatures of the galactic interaction. Such clusters are rarely observed in the other non-interacting galaxies. Aiming to identify the formation mechanism of the young massive clusters in the overlap region where the Antennae galaxies are merging, we reanalyzed ALMA Cycle 0 and Cycle 4 {data} and applied {a} prescription for cloud-cloud collision to the CO data. The main conclusions of the present study are summarized below.

\begin{enumerate}
 \item In the present analysis, we identified possible signatures for cloud-cloud collisions, the bridge features and complementary distributions, at $\sim$50 pc resolution between the two velocity components which are visible as double {peaked CO} profiles in the three cloud complexes SGMC 1, 2, and 4--5 in the overlap region. The collision signatures are identified by the method which was developed for cloud-cloud collision in the Milky Way (\cite{2018ApJ...859..166F}). We present a scenario that cloud-cloud collisions are taking pace to trigger the formation of the SSCs and SSC candidate in the Antennae. The complementary distributions are used {to estimate that the} relative displacements {are generally} $\sim$100 pc between the two clouds, which are interpreted as caused by a significant inclination angle of the relative motion to the line of sight. If we take a ratio between the displacement and the velocity, the typical time scale of the collisions is calculated to be $\sim$1 Myr{, which} is {on the same order of magnitude as} the cluster age estimated in the previous works. The results suggest that collisions between CO clouds are frequent in the overlap region, where the relative velocity $\sim$100 km s$^{-1}$ among the clouds is driven by the galactic interaction. 

\item The 1st and 2nd moment distributions are derived in the three cloud complexes and are compared with the SSCs  and {a previously identified proto-SSC candidate}. We find that the SSCs and their candidates tend to be located toward the boundaries between the two velocity clouds. The CO emission toward the SSCs is broad with a velocity span as large as 200--300 km s$^{-1}$, which was interpreted previously as the gas acceleration by the stellar feedback. The present collision interpretation suggests an alternative that the large velocity span reflects the collisional interaction between the two clouds whose velocity separation is $\sim$100 km s$^{-1}$. We present estimates that the gas motion cannot be explained by the stellar feedback if we employ the typical supernova kinetic energy as the energy source. The cloud distributions sometimes show circular shape, which is not spherical-shell like but is ring-like, {so shows no sign of expansion.} This does not lend support for the stellar feedback, {it instead} is consistent with the cavity created by cloud-cloud collision as shown by numerical simulations of cloud-cloud collision.
 
\item Two notable cases of collisions are found toward the SSC B1 and a proto-cluster candidate the Firecracker. In both of these two regions we find complementary distributions between the two clouds at {50 pc resolution for B1 and 15 pc resolution for {the} Firecracker.} {In both,} a small cloud is associated with a cavity in a larger cloud. The cavities have size similar to the small clouds, and the two clouds are linked by the bridge features. In B1, the mass of the small cloud is $\sim$5$\times$10$^7$ $M_{\odot}$, which is about 50 times larger than the cluster mass, within a cloud size of $\sim$300 pc. The collision time scale is roughly estimated to be 0.6$^{+0.6}_{-0.3}$ Myrs (=63 pc/110 km s$^{-1}$). In the Firecracker, the molecular mass of the small cloud is 9$\times$10$^6$ $M_{\odot}$ within {a} 74 pc {diameter}, several times smaller than that in B1, suggesting that the cluster possibly forming in the Firecracker {will have a} mass smaller than B1. The collision time scale here is estimated to be 0.3 Myr (=30 pc/110 km s$^{-1}$), shorter than in B1, consistent with no luminous cluster {being observed yet}. In the two regions the {colliding} velocity and average cloud density were used to estimate the pressure to be $\sim$10$^8$--10$^9$/$k_{\rm B}$ {K cm$^{-3}$}, which is consistent with the theoretically expected ambient pressure for a SSC.

\item The location of the forming clusters is probably on one side of the small cloud which is strongly interacting with the other cloud, since the collision is highly directive.  

A comparison with {Westerlund~2 and R136} with mass of 10$^4$ $M_{\odot}$ and 10$^5$ $M_{\odot}$ suggests that the sequence of cluster mass {could be understood as due} to density and collision velocity which produce high ambient pressure. Future extensive numerical simulations of colliding gas flows are required to elucidate details of cluster formation in order to provide a linkage to the ancient globulars.
\end{enumerate}
\begin{ack}
{The authors would like to thank Bradley C. Whitmore for providing ALMA Cycle 0 data treated in this paper.} This paper makes use of the following ALMA data: ADS/ JAO.ALMA \#2011.0.00876, and \#2016.1.00924.S . ALMA is a partnership of the ESO, NSF, NINS, NRC, MOST, and ASIAA. The Joint ALMA Observatory is operated by the ESO, AUI/NRAO, and NAOJ. {This paper is also based on observations made with the NASA/ESA Hubble Space Telescope, obtained from the data archive at the Space Telescope Science Institute. STScI is operated by the Association of Universities for Research in Astronomy, Inc. under NASA contract NAS 5-26555.} This study was financially supported by JSPS KAKENHI ({Grant Numbers 15H05694 and 18K13582}). This work was also financially supported by Career Development Project for Researchers of Allied Universities{, and NAOJ ALMA Scientific Research Grant Numbers 2016-03B. {K. Tsuge was supported by the ALMA Japan Research Grant of NAOJ ALMA Project, NAOJ-ALMA-232. This research is supported by NSF grants 1413231 and 1716335 (PI: K.~Johnson). This material is based upon work supported by the National Science Foundation Graduate Research Fellowship Program under Grant No. 1842490. Any opinions, findings, and conclusions or recommendations expressed in this material are those of the author(s) and do not necessarily reflect the views of the National Science Foundation.}}
\end{ack}


\renewcommand{\thefootnote}{\fnsymbol{footnote}}

\onecolumn
\begin{deluxetable}{cccccccc}
\tablewidth{14cm}
\tablecaption{{Physical properties of SGMCs}}
\label{tab:comparison1}
\tablehead{\multicolumn{2}{c}{Object} &$V_{\rm peak}$ & $T_{\rm peak}$  & d$V$ & $R$ & $M_{\rm vir}$ & $X_{\rm CO}$ mass \\
\multicolumn{2}{c}{\ }&{[}km s$^{-1}${]} &{[}K{]}&{[}km s$^{-1}${]} & {[}pc{]} &  {[}10$^{7}$ $M_{\odot}${]} & {[}10$^{7}$$M_{\odot}${]}\\
\multicolumn{2}{c}{(1)}&(2)&(3)&(4)&(5)&(6)&(7)}
\startdata
SGMC 1 & blue& 1430& 12.3 & 45&171&7 &5$^{+2}_{-1}$\\
\ & {green}&1520&  7.6 &70  & 253 & 24 &15$^{+5}_{-3}$\\ \hline
SGMC 2 &  {green}& 1515& 8.0 & 95  &250  &46  &24$^{+12}_{-6}$ \\
\ & red& 1635&4.2  & 75 &173  &19  &6$^{+3}_{-2}$ \\ \hline
SGMC 4--5 & {green}& 1490& 12.6 & 50  &158  &8  &6$^{+2}_{-1}$ \\
\ & red& 1605&  4.5&110  & 198 &49  & 4$\pm$1 \\  \hline
{The} Firecracker &  {green}& 1515& 23.6& 80  &37  &5  &2$\pm$1\\
\ & red& 1630&  9.9&60  & 61 &4.4  & 2$\pm$1
\enddata
\vspace*{-0.8cm}
\tablecomments{Column (1): Object name. Column (2): Peak velocity. Column (3): Peak intensity. Column (4):  d$V$ is a velocity dispersion. Column (5): Cloud size is defined as an effective radius =($A$/$\pi$)$^{0.5}$, where $A$ is the region enclosed by a contour of 20\% of peak integrated intensity. Column (6): Virial mass derived by using velocity dispersion. Column (7): $X_{\rm CO}$ mass is derived by using the relationship between the molecular hydrogen column density $N$(H$_{2}$) and the $^{12}$CO(1--0) intensity $W(^{12}$CO), $N$(H$_{2}$)=0.6$\times$10$^{20}$ [$W(^{12}$CO)/(K km s$^{-1}$)] (cm$^{-2}$) (\cite{2003ApJ...588..243Z,2014ApJ...795..174K}). {The value of $X_{\rm CO}$ possibly changes by a factor of three because of the difference of the relative abundance of CO to H$_{2}$. } We adopted the $^{12}$CO (3--2) to $^{\rm 12}$CO (1--0) ratios of SGMC~1, SGMC~2, and SGMC~4--5 are 0.4$\pm$0.1, 0.3$\pm$0.1, and 0.5$\pm$0.1, respectively(\cite{2012ApJ...745...65U}). {The error of the $X_{\rm CO}$ mass is estimated by using rms of CO emission and uncertainty of the $^{12}$CO (3--2) to $^{\rm 12}$CO (1--0) ratio.}}
\end{deluxetable}

\begin{deluxetable}{cccccc}
\rotate
\tablewidth{18cm}
\tablecaption{{Physical properties of collision}}
\label{tab:3}
\tablehead{\multicolumn{1}{c}{Object} &Velocity ranges of &Velocity range of& Time scale of & Evolutionary & Age of  \\
\multicolumn{1}{c}{\ }&colliding clouds&brideg features & the collision&  stage of& stellar cluster\\
\multicolumn{1}{c}{\ }&{[}km s$^{-1}${]} &{[}km s$^{-1}${]} & {[}Myr{]} & SGMCs &{[}Myr{]}  \\
\multicolumn{1}{c}{(1)}&(2)&(3)&(4)&(5)&(6)}
\startdata
SGMC 1 & 1350--1460& 1460--1505& 0.3--1.2 & Stage 1\tablenotemark{\dagger \dagger}&D2: 5.4\\
north & 1505--1550&\ &  \  & Stage 4 (D2): 3--10 Myr   \\ \hline
SGMC 1 & 1350--1460& \ & \  &Stage 2 (2 regions): 0.1--1 Myr&D: 3.9/1.45 \\
south& 1505--1550& 1460--1505& 0.45--1.8  &Stage 3 (D): 1--3 Myr&D1: 6.1 \\
\ & \ &\ &  \  &Stage 4 (D1): 3--10 Myr  & \  \\ \hline
SGMC 2 & 1400--1530&1530--1580& ---\tablenotemark{\dagger} & Stage 2 (1 region): 0.1--1 Myr  &C: 5.7/4.8  \\
\ & 1580--1700& \ &\   & Stage 3 (2 regions): 1--3 Myr &\    \\ \hline
SGMC 4--5 & 1450--1535& 1535--1560&0.3--1.2&Stage 1\tablenotemark{\dagger \dagger} & B1: 3.5/1     \\
\ & 1560--1700& \ & \ &\   & \    \\  \hline
{The} Firecracker & 1370--1510& 1510--1595& $\sim$0.3& Stage 1\tablenotemark{\dagger \dagger}  &---  \\
\ & 1595--1745&\ &  \ &\   & \   
\enddata
\vspace*{-0.8cm}
\tablecomments{Column (1): Object name. Column (2): Integrated velocity ranges of the two velocity components as shown in Section 3.1. Column (3): Velocity range of the bridge features. Column (4): Time scale of the collision calculated by a ratio of the displacement and the relative velocity of two velocity components shown in Section 5.1. Column (5):  Evolutionary stage of SGMCs defined by Whitmore et al. (2014). Also see Section 5.2, Figure 10, 11, and 12 in the present paper. Column (6): Age of SSCs same as in Table 1.}
\tablenotetext{\dagger}{We could not estimate time scale because there is no displacement between the two clouds.}
\tablenotetext{\dagger\dagger}{stage 1 is defined as compact CO emission with no radio or optical/near-IR emission. }
\end{deluxetable}

\begin{deluxetable}{clcccccccccc}
\rotate
\tablewidth{20cm}
\tablecaption{{Comparison of SSCs in the Antennae Galaxies, the Milly Way, and the Large Magellanic Cloud.}}
\label{tab:comparison2}
\tablehead{\multicolumn{2}{c}{Object} &\rm Colliding & $Colliding$  & Density of & d$V$ & $P/k_{\rm B}$& Column& Total stellar\tablenotemark{\dagger}  &Ref.\\
\multicolumn{2}{c}{\ }&gas&velocity&colliding clouds &  &  & density&mass&&\\
\multicolumn{2}{c}{\ }&&{[}km s$^{-1}${]} &$n$ {[}cm$^{-3}${]}&{[}km s$^{-1}${]} &  {[}10$^{5}$ K cm$^{-3}${]}& {[}10$^{22}$ cm$^{-2}${]}& {[}10$^{5}$ $M_{\odot}${]}&\\
\multicolumn{2}{c}{(1)}&(2)&(3)&(4)&(5)&(6)&(7)&(8)&(9)}
\startdata
{The} Firecracker & green& CO& 115 & 3400$^{+1700}_{-700}$&80&2.7$^{+1.4}_{-0.7}$$\times$10$^{4}$ &24&{--}&{--}\\
\ & red&&   &680$^{+320}_{-170}$  & 60 & 5.4$^{+2.8}_{-1.4}$$\times$10$^3$ &&&[2]\tablenotemark{b}\\ \hline
B1 &  {green}& CO& 110 & 150$\pm$30  &50  &1.1$^{+0.3}_{-0.2}$$\times$10$^3$ &10&42$\pm$8&[1][3]\tablenotemark{a} \\
\ & red& \ &\   & 50$\pm$10 &110 &340$^{+80}_{-60}$  &\ &\ & [2]\tablenotemark{b}\\ \hline
C & {green}& CO& 120 & 210$^{+100}_{-50}$  &55  &1.5$^{+0.8}_{-0.4}$$\times$10$^3$&5&41$\pm$8&[1][3]\tablenotemark{a} \\
\ & red& \ &  \ & 90$^{+50}_{-20}$ & 61&700$^{+300}_{-200}$  & \ &\ &[2]\tablenotemark{b}\\  \hline
D & {blue}& CO& 110 & 170$^{+60}_{-30}$ &46  &1.2$^{+0.4}_{-0.3}$$\times$10$^3$&10&14$\pm$2&[1][3]\tablenotemark{a} \\
\ & green& \ &  \ & 50$^{+20}_{-10}$ & 61&400$\pm$100 & \ &\ &[2]\tablenotemark{b}\\  \hline
D1 & {blue}& CO& 110 & 210$^{+60}_{-50}$  &46  &1.5$^{+0.5}_{-0.3}$$\times$10$^3$&6&16$\pm$3&[1][3]\tablenotemark{a} \\
\ & green& \ &  \ & 90$^{+50}_{-20}$ & 29&700$^{+300}_{-200}$  & \ &\ &[2]\tablenotemark{b}\\  \hline
D2 & {blue}& CO& 100 & 100$^{+40}_{-20}$  &73  &700$^{+400}_{-50}$&10&8$\pm$2&[1][3]\tablenotemark{a}\\
\ & green& \ &  \ & 110$^{+40}_{-20}$ & 51&1.1$^{+0.3}_{-0.2}$$\times$10$^3$  & \ &\ &[2]\tablenotemark{b} \\  \hline
R136 & {L-comp.}& H{\sc i}& 70 & 13.8$\pm$0.3  &45  &41$\pm$1&$\sim$1&1.2$\pm$0.3&[4]\tablenotemark{a} \\
\ & D-comp.& \ &  \ & 6.8$\pm$0.1 & 31&20$\pm$1  & \ &\ &[2]\tablenotemark{b} \\  \hline
Westerlund~2 & {blue-shift}& CO& 16 & 35$^{+12}_{-6}$  &$\sim$5 &5.5$^{+1.4}_{-1.8}$&20&0.2$\pm$0.1&[5][6]\tablenotemark{a}  \\
\ &red-shift& \ &  \ & 35$^{+12}_{-6}$ & \ &5.4$^{+1.9}_{-1.1}$  & \ &\ & [7][8]\tablenotemark{b}\\  \hline
\enddata

\vspace*{-0.5cm}
\tablecomments{Column (1): Object name. Column (2): Kind of the colliding gas. Column (3): Velocity difference between the two velocity components. Column (4): Density of colliding clouds. Column (5): Average of velocity dispersion. Column ({6}): Average of the pressure caused by a cloud-cloud collision. Column (6): The total stellar mass of the SSCs. Column (7): Maximum value of column density. Column(8): Total stellar mass of the SSCs. Column(9): References: [1] Whitmore et al. 2014, [2] the present study, [3] \citet{2007ApJ...668..168G} , [4] \citet{2015ApJ...811...76C}, [5] \citet{2007A&A...466..137A}, [6]\citet{2004ApJS..154..315W}, [7] Furukawa et al. 2009, [8] Ohama et al. 2010}

\tablenotetext{\dagger}{Relative errors in masses of \citet{2007ApJ...668..168G} are typically 20\% and dominated by photometric errors for most sources.}
\vspace*{-0.1cm}
\tablenotetext{a}{References of total stellar mass}
\vspace*{-0.1cm}
\tablenotetext{b}{Reference of physical properties of interstellar medium}
\end{deluxetable}

\twocolumn




\appendix
\section{{Results of multiscale CLEAN algorithm}}
{We confirmed how much the quality of data is improved. We made count histograms of integrated intensities from the data of \citet{2014ApJ...795..156W} and the data of this paper as shown in Figure \ref{figa1}. The integral velocity range is 1300--1700 km s $^{-1}$, and the beam size, grid size, velocity resolution, and analysis region are smoothed to be the same. The reanalyzed data (this paper) recovers CO emission (red shaded area in Figure \ref{figa1}c), and reduce negative level (blue shaded area in Figure \ref{figa1}c) more than data of \citet{2014ApJ...795..156W}.} {We shows channel maps of data. The quality of our data is improved in particular in a velocity range of 1452--1552 km $^{-1}$ as shown in Figure\ref{figa2}.}

\begin{figure*}[htbp]
\begin{center}
\includegraphics[width=14cm]{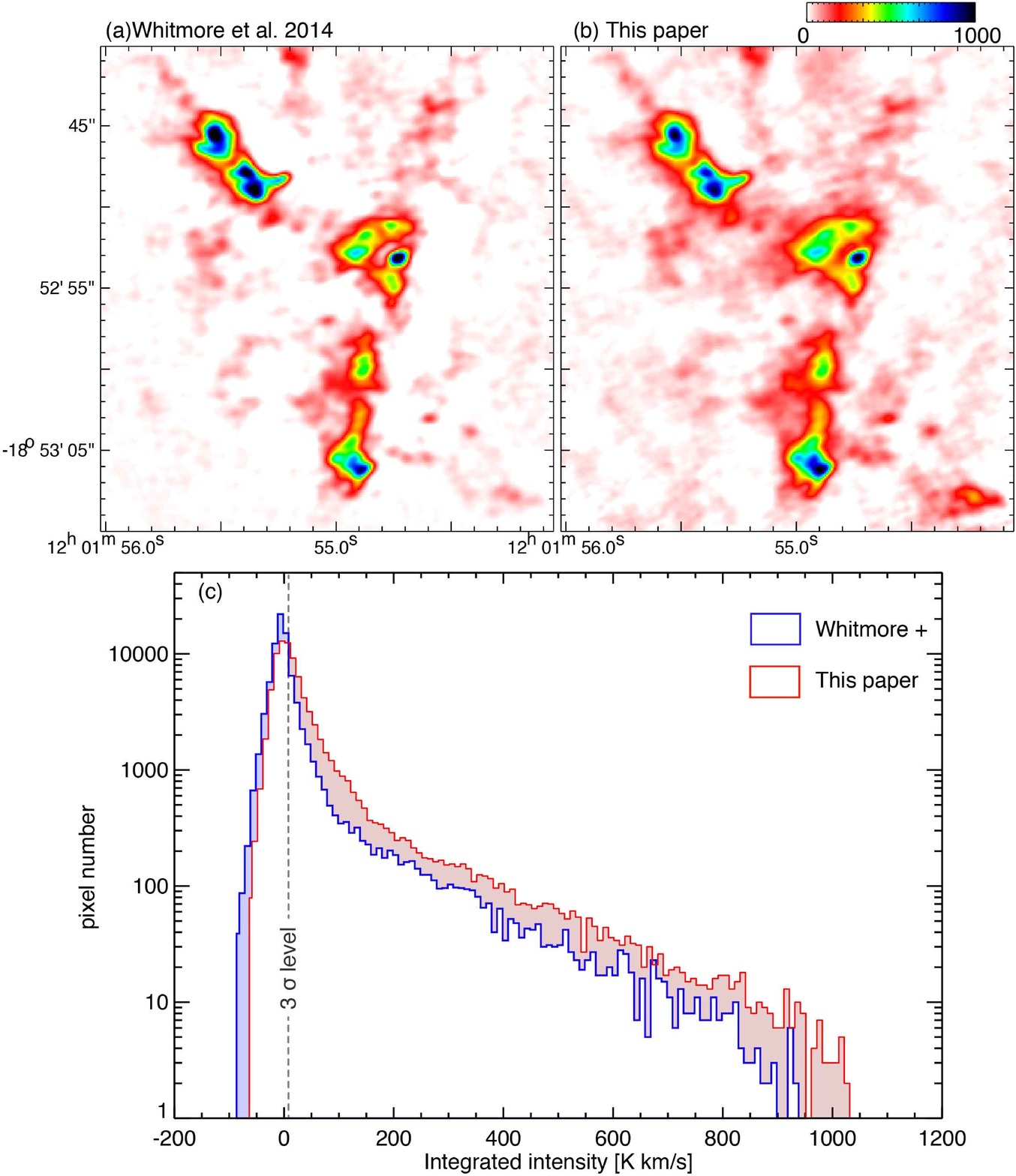}
\end{center}
\caption{(a) Total integrated intensity map of $^{12}$CO(3--2) toward the overlap region , which was analyzed by \citet{2014ApJ...795..156W}.  (b)  Same as (a), but the $^{12}$CO(3--2) data was reanalyzed by multiscale CLEAN algorithm in this paper. (c) Histogram of the number of pixel. The horizontal and vertical axes are integrated intensity and the number of pixel, respectively.} 
\label{figa1}
\end{figure*}%

\begin{figure*}[htbp]
\begin{center}
\includegraphics[width=11cm]{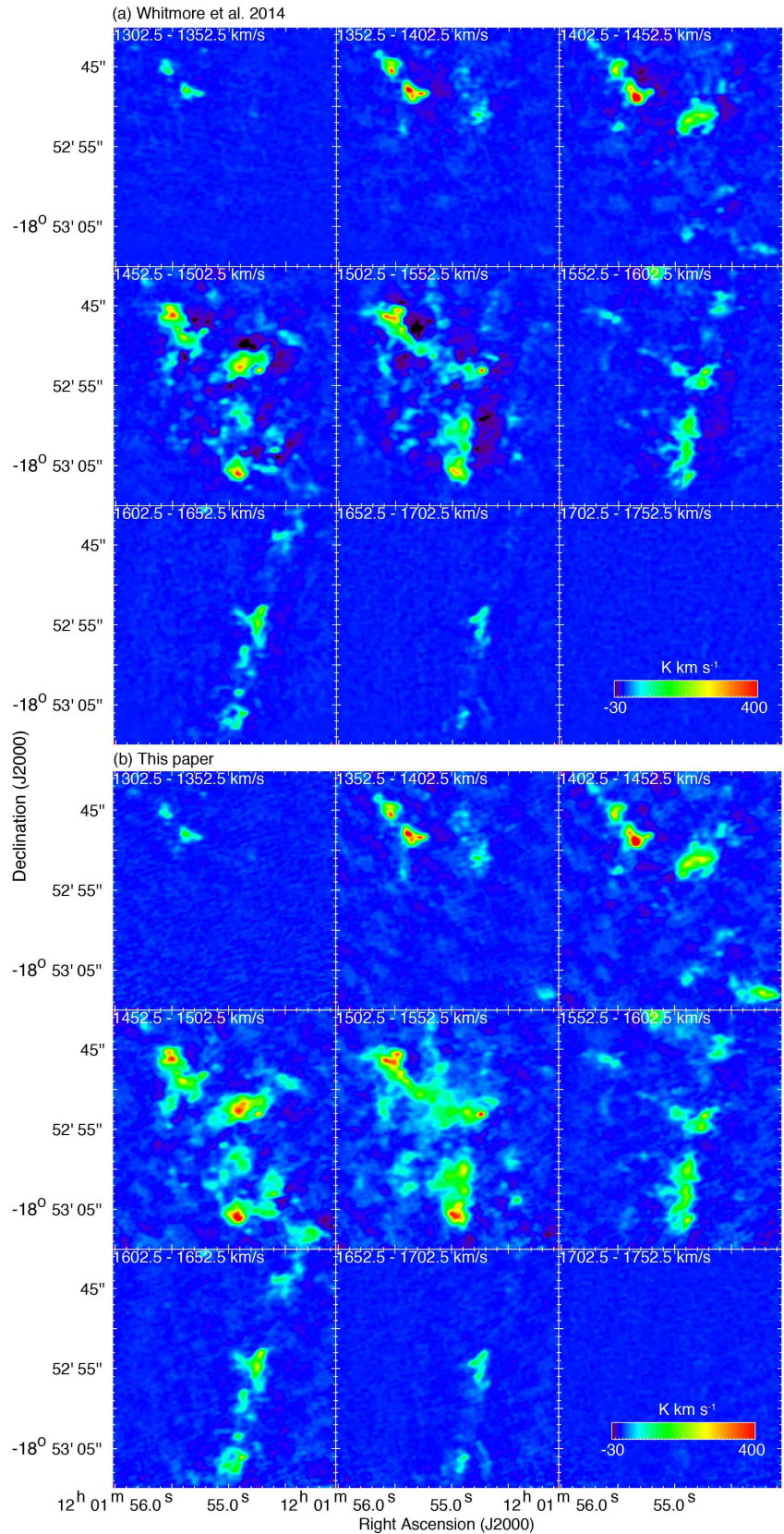}
\end{center}
\caption{(a) Velocity channel maps of $^{12}$CO(3--2) toward the overlap region with a velocity step of 50 km s$^{-1}$, which was analyzed by \citet{2014ApJ...795..156W}. (b) Same as (a), but the $^{12}$CO(3--2) data was reanalyzed by multiscale CLEAN algorithm. } 
\label{figa2}
\end{figure*}%

\clearpage

\section{CO channel maps of position velocity diagrams }
We show the 11 declination--velocity diagrams of the $^{12}$CO(3--2) toward SGMC1, SGMC2, and SGMC4--5. The integration range is $\timeform{1.0"}$ ($\sim$107 pc), and the integration range is shifted from east to west in $\timeform{1.0"}$ step

\begin{figure*}[htbp]
\begin{center}
\includegraphics[width=16cm]{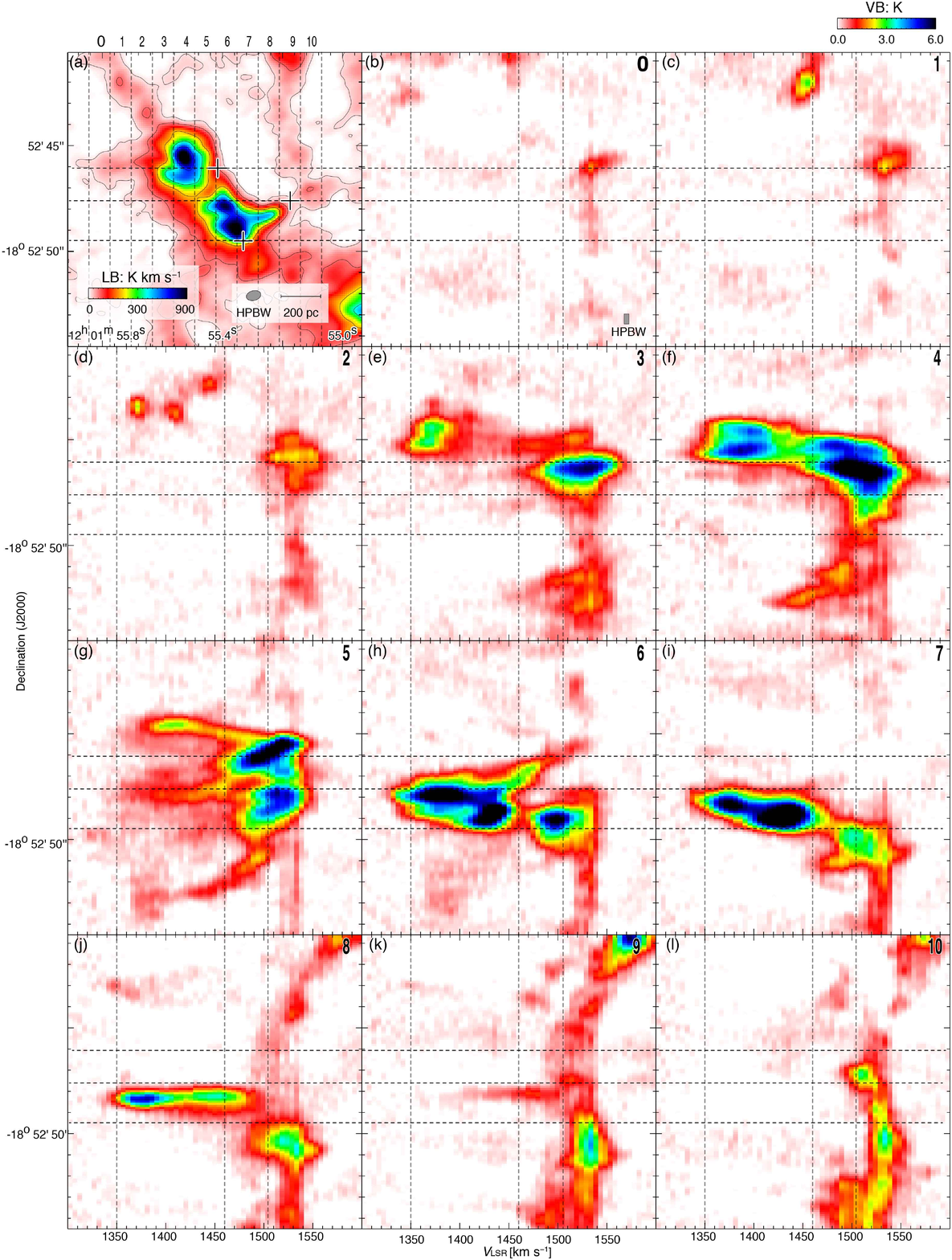}
\end{center}
\caption{Channel maps of declination--velocity diagrams for SGMC1. (a) $^{12}$CO(3--2) total integrated intensity map of SGMC1. The integration velocity range and contour levels are the same as in Figure \ref{fig6}a. Dashed lines indicate the integration ranges of declination-velocity diagrams in R.A.. (b)-(l) Declination-velocity diagrams of $^{12}$CO (3--2) . The upper right number denotes the integration range in (a). } 
\label{figa3}
\end{figure*}%

\begin{figure*}[htbp]
\begin{center}
\includegraphics[width=16cm]{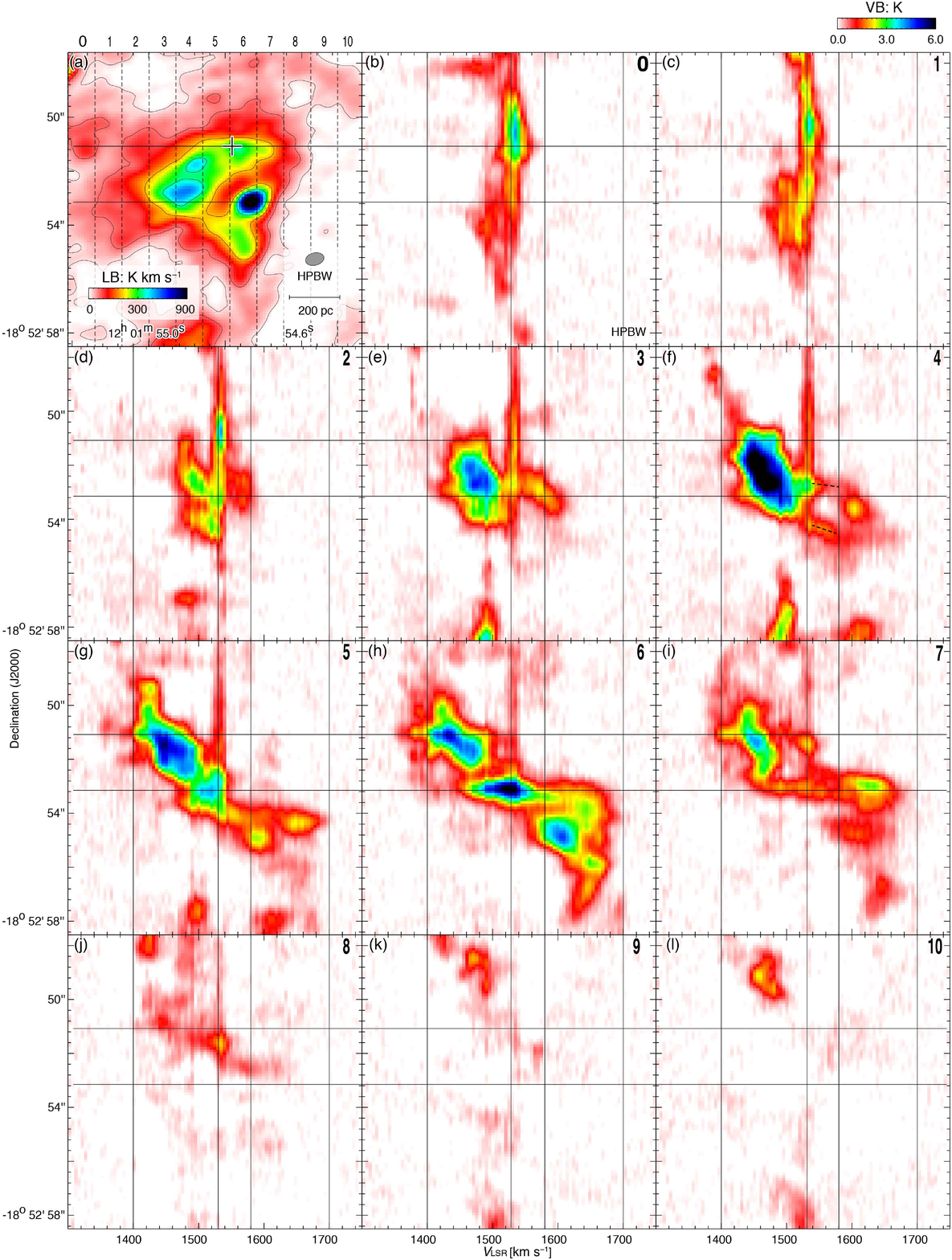}
\end{center}
\caption{Channel maps of declination--velocity diagrams for SGMC2. (a) $^{12}$CO(3--2) total integrated intensity map of SGMC2. The integration velocity range and contour levels are the same as in Figure \ref{fig7}a. Dashed lines indicate the integration ranges of declination-velocity diagrams in R.A.. (b)-(l) Declination-velocity diagrams of $^{12}$CO (3--2) . The upper right number denotes the integration range in (a). } 
\label{figa4}
\end{figure*}%

\begin{figure*}[htbp]
\begin{center}
\includegraphics[width=16cm]{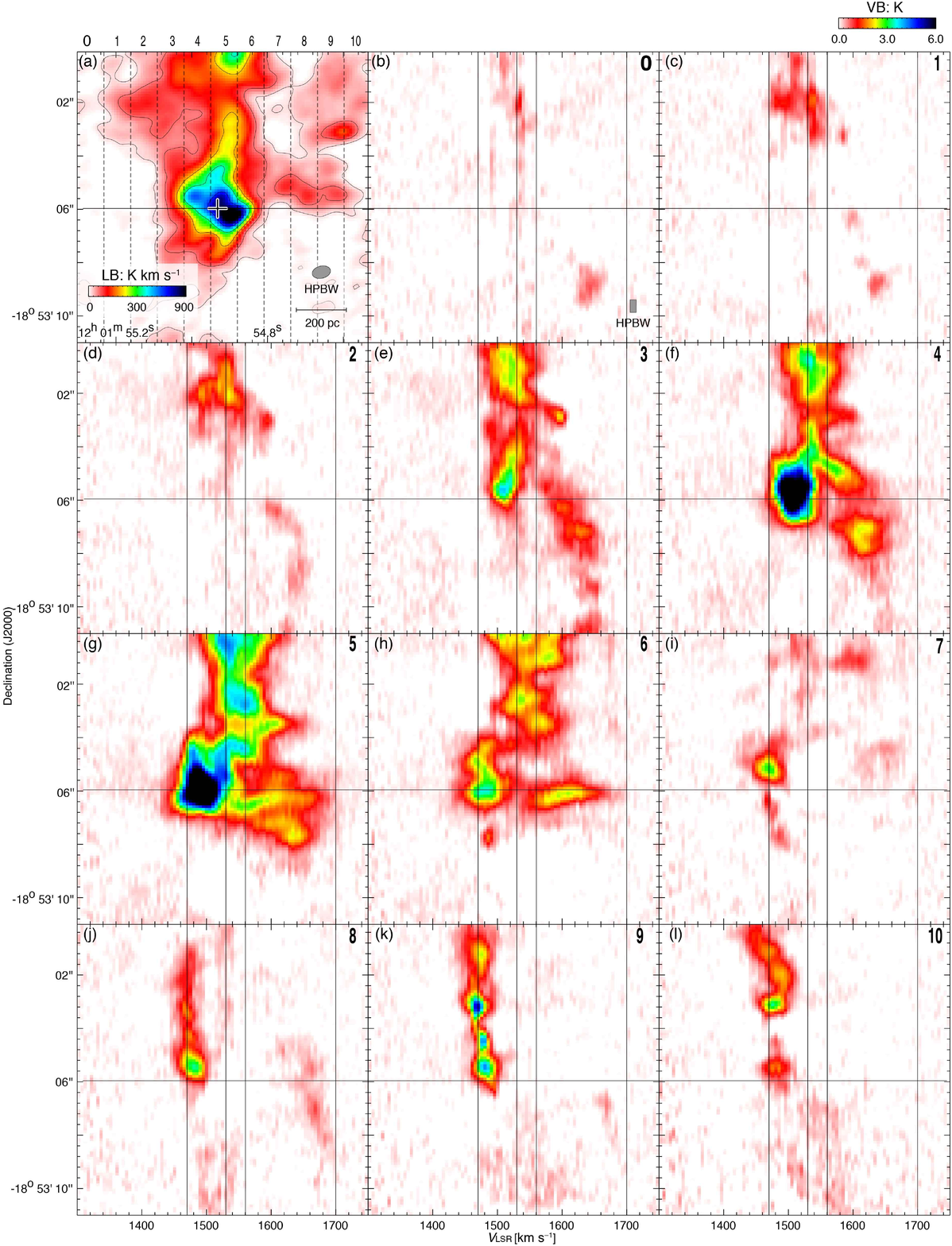}
\end{center}
\caption{Channel maps of declination--velocity diagrams for SGMC4--5. (a) $^{12}$CO(3--2) total integrated intensity map of SGMC4--5. The integration velocity range and contour levels are the same as in Figure \ref{fig8}a. Dashed lines indicate the integration ranges of declination-velocity diagrams in R.A.. (b)-(l) Declination-velocity diagrams of $^{12}$CO (3--2) . The upper right number denotes the integration range in (a). } 
\label{figa5}
\end{figure*}%

\end{document}